\documentclass[prb,reprint,showpacs,floatfix]{revtex4-1}
\usepackage[pdftex]{graphicx} 
\usepackage{dcolumn} 
\usepackage{bm} 
\usepackage{amssymb} 
\usepackage{amsmath} 
\usepackage{hyperref}
\usepackage{footmisc}
\usepackage[latin1]{inputenc}
\usepackage{tikz}
\usetikzlibrary{shapes,arrows}
\usepackage{float}
\begin{document}
\title{Local density approximations from finite systems} 
\date{\today}
\author{M.\ T.\ Entwistle}
\affiliation{Department of Physics, University of York, and European Theoretical Spectroscopy Facility, Heslington, York YO10 5DD, United Kingdom}
\affiliation{Fitzwilliam College, University of Cambridge, Cambridge CB3 0DG, United Kingdom}
\author{M.\ J.\ P.\ Hodgson}
\author{J.\ Wetherell}
\author{B.\ Longstaff}
\altaffiliation{Present Address: Department of Mathematics, Imperial College London, London SW7 2AZ, United Kingdom}
\author{J.\ D.\ Ramsden}
\author{R.\ W.\ Godby}
\affiliation{Department of Physics, University of York, and European Theoretical Spectroscopy Facility, Heslington, York YO10 5DD, United Kingdom}
\begin{abstract} 
The local density approximation (LDA) constructed through quantum Monte Carlo calculations of the homogeneous electron gas (HEG) is the most common approximation to the exchange-correlation functional in density functional theory. We introduce an alternative set of LDAs constructed from slablike systems of one, two and three electrons that resemble the HEG within a \textit{finite} region, and illustrate the concept in one dimension. Comparing with the exact densities and Kohn-Sham potentials for various test systems, we find that the LDAs give a good account of the self-interaction correction, but are less reliable when correlation is stronger or currents flow.
\end{abstract}


\maketitle

\section{Introduction} 
Density functional theory \cite{DFT} (DFT) is the most widely used method to perform ground-state electronic structure calculations of many-electron systems in condensed matter physics and many areas of materials science. In the Kohn-Sham (KS) approach \cite{KS_LDA} to DFT the real many-electron system, which is governed by the often unsolvable many-body Schr\"{o}dinger equation, is replaced by a fictitious system of noninteracting electrons with the same density. The absence of interaction allows the system to be described by several single-particle Schr\"{o}dinger equations (KS equations) in which the electrons are moving in an effective local potential $V_{\rm{KS}}$. While in principle an exact theory, in practice the accuracy of DFT depends on its ability to approximate the unknown exchange-correlation (xc) part of the KS functional \cite{KS_LDA}.

The local density approximation \cite{KS_LDA} (LDA) is the most common approximation to the xc potential $V_{\rm{xc}}$. The LDA is traditionally based on knowledge of the energy of the infinite three-dimensional (3D) homogeneous electron gas \cite{QMC_HEG} (HEG), in which the electrons are commonly viewed as delocalized. Although local approximations have had major success in many cases \cite{LDA_Success,LDA_Success2}, they fail in other situations. A notable failing is the inability to correctly cancel the spurious electron self-interaction \cite{PerdewZunger,SIC,SIC2}, an error introduced by the Hartree potential. Also, the xc potential far from a finite system decays exponentially in an LDA \cite{PerdewZunger,CoulombDecay/ExpDecay}, rather than following the Coulomb-like $-1/r$ decay present in the exact $V_{\rm{xc}}$ \cite{CoulombDecay,CoulombDecay/ExpDecay}. These failings lead to errors in the KS orbitals \cite{Errors_Orbitals}. Many time-dependent DFT \cite{TDDFT,TDDFT2} (TDDFT) calculations are performed by applying the LDA adiabatically (ALDA), which further ignores the dependence of $V_{\rm{xc}}$ on a system's history and initial state, focusing instead on the instantaneous electron density.  Local approximations are known to break down in a number of cases\cite{ALDA,ALDA2,ALDA3,ALDA4,ALDA5,ALDA6,ALDA7,ALDA8,ALDA9,QuantumTransport1,QuantumTransport2}, in particular where there is strong correlation in ground-state systems and/or strong current flow when extended to time-dependent systems.

In this paper we introduce a set of LDAs constructed from systems of one, two and three electrons. In contrast to a conventional LDA which is constructed through accurate (but not exact) quantum Monte Carlo (QMC) simulations of the HEG approaching the thermodynamic limit \cite{QMC_HEG}, our approach is to obtain a set of LDAs constructed from exact \textit{finite} systems resembling the HEG. We refer to these finite systems as `slabs' to emphasize that the electron density is dominated by a region of homogeneity, but decays exponentially to zero near the edges. We compare these LDAs with one another and with conventional HEG-based LDAs. We illustrate our approach in one dimension (1D), complementing other 1D LDAs that have been constructed through QMC calculations, either with a softened Coulomb interaction \cite{LDA_QMC} or a specified transverse confining potential \cite{LDA_INT,LDA_INT2}, or through other approaches \cite{LDA_DMRG,LDA_AdHoc3}.

We employ our iDEA code \cite{iDEA} which determines the exact, fully-correlated, many-body wave function for a finite system of electrons interacting via the appropriately softened Coulomb repulsion \cite{SoftenedCoulomb} $(|x-x'|+1)^{-1}$. We then find the corresponding exact KS system through our reverse-engineering algorithm \cite{iDEA}. The electrons are treated as \textit{spinless} to more closely approach the nature of exchange and correlation in many-electron systems\footnote{Spinless electrons obey the Pauli principle but are restricted to a single spin type. Systems of two or three spinless electrons exhibit features that would need a larger number of spin-half electrons to become apparent. For example, two spinless electrons experience the exchange effect, which is not the case for two spin-half electrons in an $S = 0$ state. Furthermore, spinless KS electrons occupy a greater number of KS orbitals.}. We then apply the LDAs to a variety of ground-state systems and find that they yield accurate densities for systems dominated by either the exchange energy or by the self-interaction correction. We demonstrate that the LDAs break down as correlation becomes strong, including when applied adiabatically to a time-dependent system. 

\section{Constructing the LDAs} 
\subsection{The finite model homogeneous systems} 
We choose a set of \textit{finite} locally homogeneous systems in order to replicate the HEG from which traditional LDAs are usually constructed. To generate these slab systems we use our optimization code which finds the correct external potential $V_{\rm{ext}}$ for a target system with a desired electron density $n_{\rm{T}}(x)$. After making an initial guess for the system, the exact many-body wave function is calculated and $V_{\rm{ext}}$ is refined iteratively, following the method used for the KS potential in Ref.~\onlinecite{iDEA}. 

\begin{figure}[htbp] 
\centering
\includegraphics[width=1.0\linewidth]{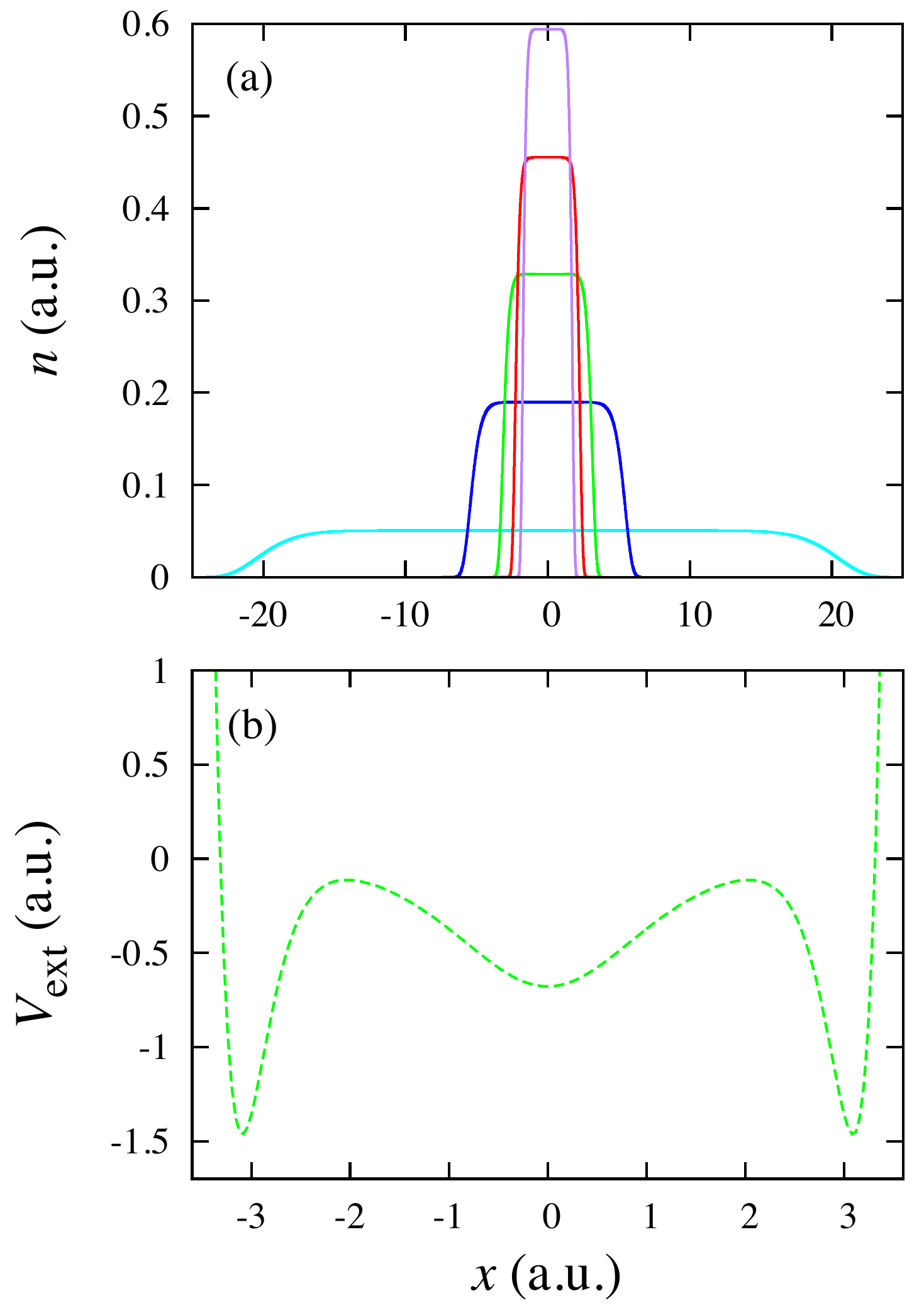}
\caption{(a) The exact many-body electron density (solid lines) for a selection of the two-electron slab systems. The density is of the form $n(x) = n_{0}\exp\big[-10^{-11}(mx)^{12}\big]$, to generate a uniform plateau region that decays exponentially at the edges. (b) The optimized external potential (dashed green line) for a typical two-electron slab system [middle density in (a), $n_{\rm{0}} \approx 0.33$]. }
\label{Slab2e}
\end{figure}

The slab systems are chosen such that the majority of the density is approximately uniform over a plateau region of value $n_{0}$ with the edges of the system decaying rapidly to zero [Fig.~\ref{Slab2e}(a)]. We therefore choose a target density of the form $n_{\rm{T}}(x) = n_{\rm{0}}\exp\big[-10^{-11}(mx)^{12}\big]$, where $m$ is a scaling factor chosen so that the density integrates to the appropriate number of electrons (2 or 3). The external potential required to obtain the desired density profile has a non-trivial spatial dependence [Fig.~\ref{Slab2e}(b)]. Sets are created from both two and three-electron slab systems and the densities cover a typical range (up to 0.6 a.u.\footnote{We use Hartree atomic units: $m_{\rm{e}} = \hbar = e = 4\pi \varepsilon_{0} = 1$}) that will be encountered when the LDAs are applied to test systems.

\subsection{Generating the LDAs} 

Having characterized the many-electron slab systems we then find the corresponding KS systems through our reverse-engineering code. By calculating the exact xc energy $E_{\rm{xc}}$ for each slab system we obtain a set of data points for the exact xc energy per electron $\varepsilon_{\rm{xc}} = E_{\rm{xc}}/N$ in terms of the electron density of the plateau regions, i.e., at this stage neglecting the inhomogeneous regions of the slab systems. We then apply a fit to determine a functional form of $\varepsilon_{\mathrm{xc}}(n)$ for the two-electron ($2e$) (shown in Fig.~\ref{exc2e}) and three-electron ($3e$) slab systems \footnote{See Supplemental Material for the initial LDAs.}. These initial LDAs are refined below.

\begin{figure}[htbp] 
\centering
\includegraphics[width=1.0\linewidth]{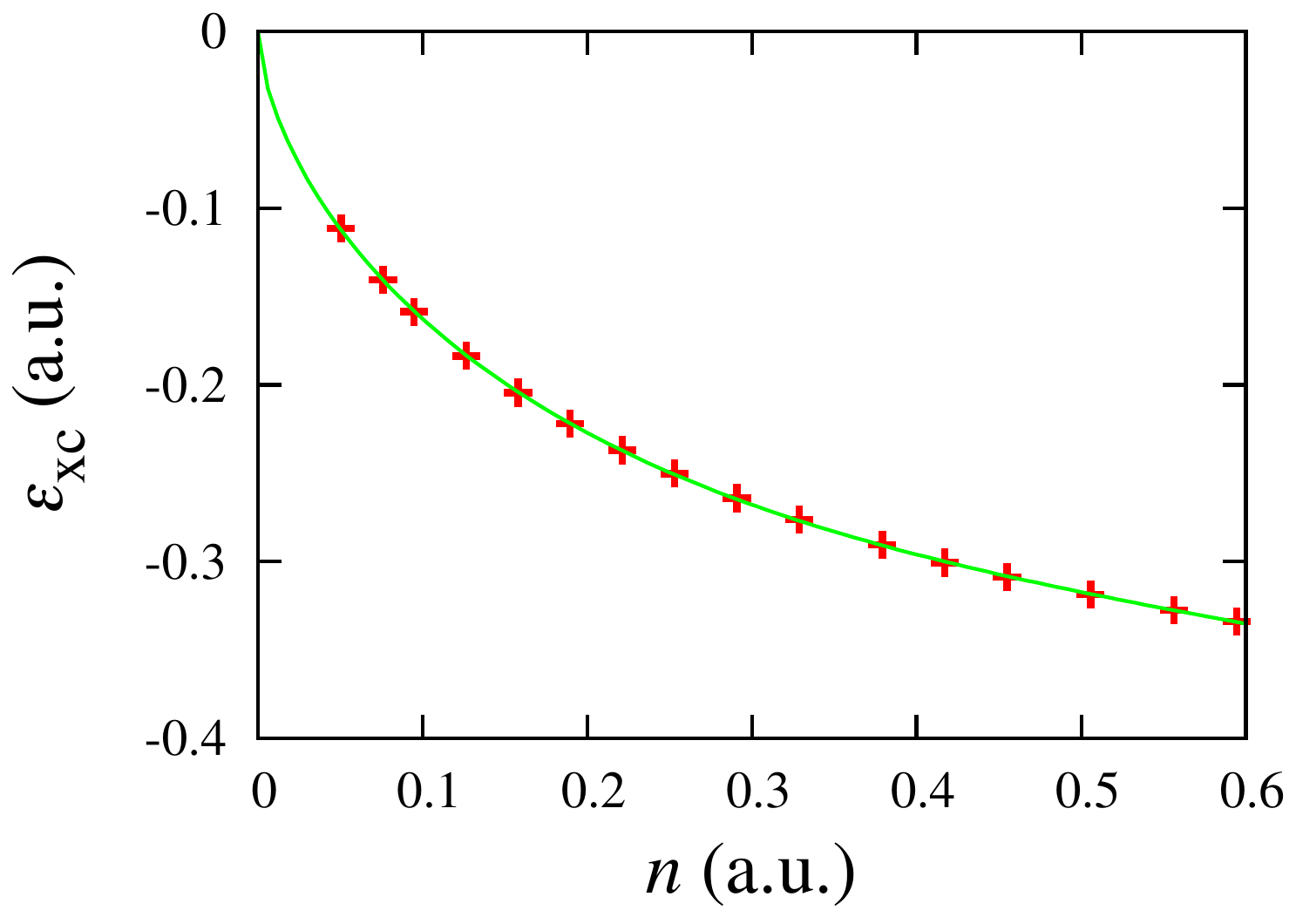}
\caption{The exact $\varepsilon_{\rm{xc}}$ (red crosses) for the $2e$ slab systems with the assigned values for the electron density being that of the plateau region $n_{\rm{0}}$. The fit applied (solid green line) is of the form $\varepsilon_{\rm{xc}} = (A + Bn + Cn^{2})n^{D}$, where $A, B, C, $ and $ D$ are constants. This initial LDA is subsequently refined by applying it to the slabs themselves (see text).}
\label{exc2e}
\end{figure}

To approximate the xc energy of an inhomogeneous system the LDA focuses on the local electron density at each point in the system:
\begin{equation}
E_{\rm{xc}}^{\rm{LDA}}[n] = \int n(x)\varepsilon_{\rm{xc}}(n)dx \label{E_xc},
\end{equation}
where $\varepsilon_{\rm{xc}}(n)$ is the xc energy per electron of a HEG of density $n$ in a traditional LDA. This approximation becomes exact in the limit of the HEG, i.e., the systems from which an LDA is constructed. In the same spirit, we require our LDAs that have been constructed from finite slab systems to yield the exact xc energies when applied to those same slab systems. 

We apply the initial LDAs to the $2e$ and $3e$ slab systems. Small errors in the xc energy $\Delta E_{\rm{xc}}$ are found due to the inhomogeneous regions of the slab systems being ignored when the LDAs were originally constructed. We use the calculated errors to determine refined forms for $\varepsilon_{\rm{xc}}$ in the LDAs\footnote{See Supplemental Material for the errors.}, $\varepsilon_{\rm{xc}}(n) \rightarrow \varepsilon_{\rm{xc}}(n) - \Delta E_{\rm{xc}}(n)/N$:

\begin{equation}
\begin{split}
2e: \varepsilon_{\rm{xc}}(n) = (-0.74 + 0.68n - 0.38n^{2})n^{0.604}
\label{2e_exc_refined} \\
\end{split}
\end{equation}
\begin{equation}
\begin{split}
3e: \varepsilon_{\rm{xc}}(n) = (-0.77 + 0.79n - 0.48n^{2})n^{0.61}.
\label{3e_exc_refined} \\
\end{split}
\end{equation}
These refined forms for $\varepsilon_{\rm{xc}}$ reduce $\Delta E_{\rm{xc}}$ from $2\%-3\%$ to below $0.5\%$ when applied to the slab systems. This refinement process is thus determined to be sufficient. 

When the LDAs are applied to inhomogeneous systems it is the xc potential that is the crucial quantity used to determine the electron density. $V_{\mathrm{xc}}$ is the functional derivative of the xc energy which in the LDA becomes
\begin{equation}
V_{\rm{xc}}^{\rm{LDA}}(x) = \frac{\delta E_{\rm{xc}}^{\rm{LDA}}[n]}{\delta n(x)} = \varepsilon_{\rm{xc}}(n(x)) + n(x)\frac{d\varepsilon_{\rm{xc}}}{dn}\bigg|_{n(x)} \label{V_xc}.
\end{equation}
The following expressions are therefore obtained from Eq.~(\ref{2e_exc_refined}) and Eq.~(\ref{3e_exc_refined}), respectively:
\begin{equation}
\begin{split}
2e: V_{\rm{xc}}(n) = (-1.19 + 1.77n - 1.37n^{2})n^{0.604}
\label{2e_Vxc_refined} \\
\end{split}
\end{equation}
\begin{equation}
\begin{split}
3e: V_{\rm{xc}}(n) = \ & (-1.24 + 2.1n - 1.7n^{2})n^{0.61}.
\label{3e_Vxc_refined} \\
\end{split}
\end{equation}

\subsection{An LDA from \textit{one-electron} slabs} 
So far we have constructed LDAs from systems of two and three interacting electrons. Owing to the absence of the Coulomb interaction it is simple to construct \textit{one-electron} ($1e$) slab systems. In a $1e$ system the Hartree energy is entirely self-interaction and so the xc energy is entirely self-interaction correction:
\begin{equation}
\varepsilon_{\rm{xc}} = E_{\rm{xc}} = -E_{\rm{H}} = -\frac{1}{2}\int\int \frac{n(x)n(x')}{|x-x'|+1}dxdx', \label{1e_exc}
\end{equation}
where the electron density is of the same form as the $2e$ and $3e$ slab systems, $n(x) = n_{0}\exp\big[-10^{-11}(mx)^{12}\big]$.

A selection of slab systems is chosen and $\varepsilon_{\rm{xc}}$ is calculated to build up a set of data points. An initial fit is found \footnote{See Supplemental Material for the initial $1e$ LDA.} and the same refinement process used in the $2e$ and $3e$ slab systems is applied. From this an expression for $\varepsilon_{\rm{xc}}$ and $V_{\rm{xc}}$ follows:
\begin{equation}
\begin{split}
1e: \varepsilon_{\rm{xc}}(n) = (-0.803 + 0.82n - 0.47n^{2})n^{0.638}
\label{1e_exc_refined} \\
\end{split}
\end{equation}
\begin{equation}
\begin{split}
1e: V_{\rm{xc}}(n) = \ & (-1.315 + 2.16n - 1.71n^{2})n^{0.638}.
\label{1e_vxc_refined} \\
\end{split}
\end{equation}

\subsection{Comparison of 1e, 2e and 3e LDAs}
\begin{figure}[htbp] 
\centering
\includegraphics[width=1.0\linewidth]{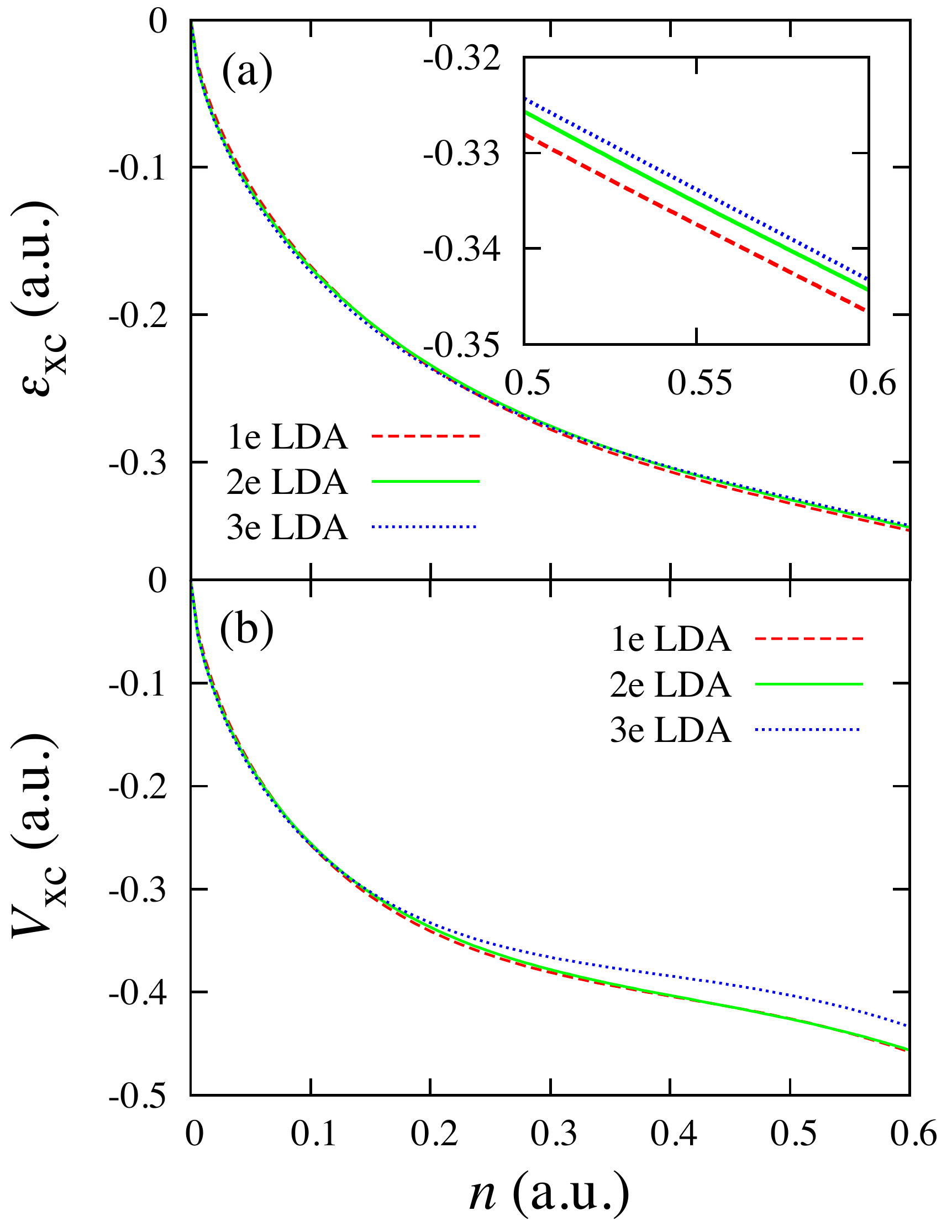}
\caption{(a) The refined curves for $\varepsilon_{\rm{xc}}$ in the $1e$ (dashed red line in both plots), $2e$ (solid green line in both plots), and $3e$ (dotted blue line in both plots) LDAs. Inset: Close-up of the three curves at higher densities. The slight deviations at higher densities are numerically significant. There is a clear progression from $1e$ to $2e$ to $3e$. (b) The refined curves for $V_{\rm{xc}}$ in the $1e$, $2e,$ and $3e$ LDAs. The closeness of the three curves, in each case, is striking.}
\label{1e2e3e}
\end{figure}

We now compare the $1e$, $2e$, and $3e$ LDAs that have been developed. The strong similarity between the three LDAs can be seen in the refined curves for $\varepsilon_{\rm{xc}}$ [Fig.~\ref{1e2e3e}(a)]. This is remarkable due to physical correlation being absent in one-electron systems and $\varepsilon_{\rm{xc}}$ consisting entirely of self-interaction correction. While the three curves effectively overlap at low densities, they deviate slightly at higher densities [inset of Fig.~\ref{1e2e3e}(a)] with these deviations being numerically significant. There is a clear progression from $1e$ to $2e$ to $3e$.

This is also seen in the refined curves for $V_{\rm{xc}}$ [Fig.~\ref{1e2e3e}(b)]. The $1e$ and $2e$ overlap at high densities with the $3e$ curve deviating slightly.

\subsection{The one-dimensional homogeneous electron gas} 
Various parametrizations\cite{PerdewZunger, PerdewWang, ColePerdew} of QMC calculations show that in the case of a 3D HEG, the exchange energy per electron $\varepsilon_{\rm{x}}$  is dominant over the correlation energy per electron $\varepsilon_{\rm{c}}$, particularly for higher densities. We solve the Hartree-Fock (HF) equations to determine the exact $\varepsilon_{\rm{x}}$ for a 1D HEG consisting of an infinite number of electrons interacting via the softened Coulomb repulsion $u(x-x')$:
\begin{equation}
\varepsilon_{\rm{x}} = -\frac{1}{8\pi^{2}n}\int_{-\pi n}^{\pi n}dk \int_{-\pi n}^{\pi n}dk'u(k'-k),\label{HF}
\end{equation}
where the Fourier transform of $u(x-x')$ is integrated over the plane defined by the Fermi wave vector $k_{\rm{F}}=\pi n$, for a HEG of density $n$.

\begin{figure}[htbp] 
\centering
\includegraphics[width=1.0\linewidth]{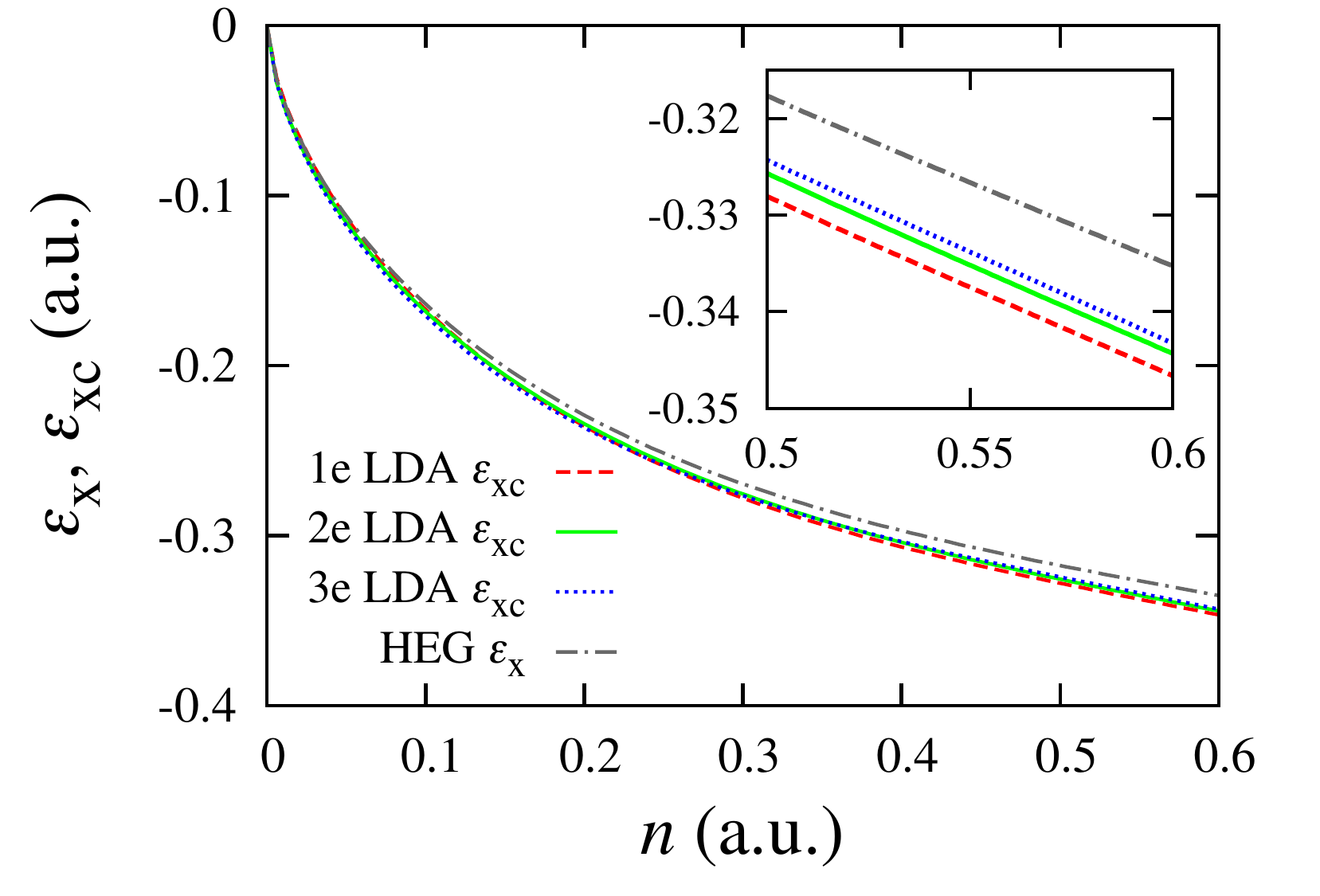} 
\caption{The exact exchange energy $\varepsilon_{\rm{x}}$ (dotted-dashed dark-gray line) of a 1D HEG of density $n$. The $\varepsilon_{\rm{xc}}$ curves in the $1e$ (dashed red line), $2e$ (solid green line), and $3e$ (dotted blue line) LDAs are repeated from Fig.~\ref{1e2e3e} for comparison. Inset: Close-up of the four curves at higher densities. All four curves are remarkably similar, indicating the importance of exchange in 1D, and showing the similarity of the different LDA approaches in 1D systems.}
\label{HEG_ex}
\end{figure}

Using Eq.~(\ref{HF}) we calculate $\varepsilon_{\rm{x}}$ for a set of HEGs covering the range of densities used in the LDAs. We then apply a fit to determine a functional form of $\varepsilon_{\rm{x}}$ for the 1D HEG. From this we find that the $\varepsilon_{\rm{x}}$ curve in the 1D HEG is surprisingly close to the $\varepsilon_{\rm{xc}}$ curves in the $1e$, $2e$, and $3e$ LDAs [Fig.~\ref{HEG_ex}]. This suggests that $\varepsilon_{\rm{x}}$ is the dominant term in $\varepsilon_{\rm{xc}}$ in the case of a 1D HEG, even more so than in the 3D case. 

In Ref.~\onlinecite{LDA_QMC}, QMC calculations of a 1D HEG of electrons interacting through a slightly different softened Coulomb interaction are used to determine a functional form for $\varepsilon_{\rm{c}}$. We evaluate $\varepsilon_{\rm{x}}$ using the method of Eq.~(\ref{HF}) for this HEG, and find $\varepsilon_{\rm{c}}$ to be of the order of a few percent of $\varepsilon_{\rm{xc}}$, except in the low-density limit. Assuming this result to be applicable to our own (very similar) 1D HEG, we conclude that the $\varepsilon_{\rm{xc}}$ curve constructed from a HEG for our softened interaction would be close to the three $\varepsilon_{\rm{xc}}$ curves for our LDAs constructed from finite systems [Fig.~\ref{HEG_ex}]. That is, in 1D, an LDA constructed from small finite systems is very similar to one constructed from the infinite HEG. 

\subsection{Extension to higher dimensions}
In Ref.~\onlinecite{LDA_Finite}, an LDA is developed that satisfies exact constraints derived from 3D finite systems, with the intention of it being more applicable to finite systems than the conventional LDA. We find the concept of constructing LDAs from 3D finite systems in their own right to be feasible. For this feasibility study we have restricted our consideration to one-electron 3D systems. By generating a set of 3D one-electron systems with a slablike radial density profile (analogous to the 1D slab systems), we develop an LDA that exhibits a form for $\varepsilon_{\rm{xc}}$ that is qualitatively similar to that of traditional local approximations constructed through QMC calculations. 

\begin{figure}[htbp] 
\centering
\includegraphics[width=1.0\linewidth]{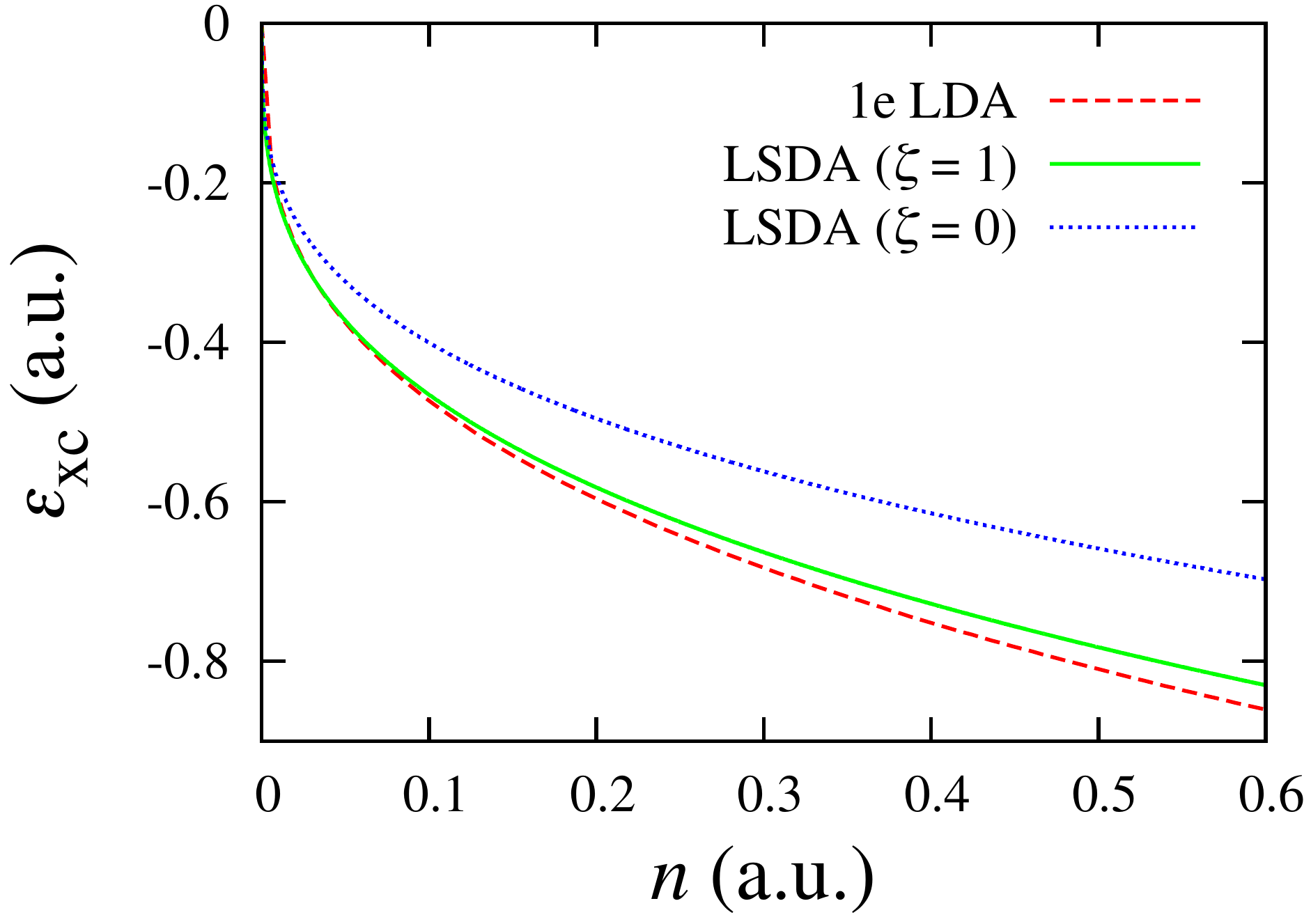} 
\caption{The $\varepsilon_{\rm{xc}}$ curve (dashed red line) for the LDA that has been developed from 3D $1e$ finite systems (which are fully spin-polarized), along with the fully spin-polarized ($\zeta = 1$) (solid green line) and fully spin-unpolarized ($\zeta = 0$) (dotted blue line) traditional LSDA~\cite{PerdewZunger}. We find the $1e$ LDA and LSDA ($\zeta = 1$) to be remarkably similar.}
\label{3D_exc}
\end{figure}

Specifically, we compare our 3D $1e$ LDA with the local spin density approximation (LSDA) as parametrized by Perdew and Zunger~\cite{PerdewZunger} [Fig.~\ref{3D_exc}]. We find that the $1e$ LDA is remarkably close to the fully spin-polarized ($\zeta = 1$) LSDA. We believe this is a fairer comparison than the fully spin-unpolarized ($\zeta = 0$) LSDA, as our 3D finite systems contain one spin-half electron; i.e., they are fully spin-polarized. \footnote{This raises the question as to whether an LDA constructed from finite systems containing two spin-half electrons in the $S = 0$ state (fully spin-unpolarized) would be much closer to the LSDA ($\zeta = 0$).} 

We compare the $1e$ LDA with the LSDA by applying them self-consistently to ground-state hydrogen and helium atoms to approximate the total energies, given in Table~\ref{Atoms_table}. We apply the ($\zeta = 0$) and ($\zeta = 1$) LSDA to both atoms (feigning the densities to be fully spin-unpolarized and fully spin-polarized, respectively, for comparison purposes). We find that our simple $1e$ LDA is able to approximate the energy in both cases, performing better than the LSDA in the case of the hydrogen atom (fully spin-polarized system), and slightly worse in the helium atom (fully spin-unpolarized system). 
\begin{table}[htb]
\caption{The exact total energies of atoms in their ground-state, along with the energies obtained by applying the $1e$ LDA and the conventional LSDA self-consistently. All energies are in a.u.}
\label{Atoms_table}
\resizebox{\columnwidth}{!}{%
\begin{ruledtabular}
\begin{tabular}{c*{5}{>{$}c<{$}}}
  Atom  & E^{\rm{exact}}         & E^{1e \ \rm{LDA}}        & E^{\rm{LSDA}(\zeta=0)}        & E^{\rm{LSDA}(\zeta=1)}             \\
  \hline
H  & -0.500 & -0.482 & -0.446 & -0.479 \\
He  & -2.90 & -3.06 & -2.83 & -3.01 
\end{tabular}
\end{ruledtabular}
}
\end{table}

\subsection{Physics of the slab systems} 
To determine what fraction of $\varepsilon_{\rm{xc}}$ for the (many-electron) slab systems is due to  $\varepsilon_{\rm{x}}$ and what fraction is due to $\varepsilon_{\rm{c}}$, we apply the HF method self-consistently to the $2e$ and $3e$ slab systems (as defined by the external potentials). We find the HF method reproduces accurate densities for high-density slab systems but breaks down for low-density slab systems. This suggests that correlation (which the HF method neglects) increases as we progress to lower densities, which is consistent with QMC calculations of the 3D HEG and other systems~\cite{QMC_Silicon}. 

In both the $2e$ and $3e$ slab systems, we calculate $\varepsilon_{\rm{x}}$ to be the dominant component in $\varepsilon_{\rm{xc}}$, with $\varepsilon_{\rm{c}}$ increasing as we move to lower density slab systems. However, we see that the correlation energy remains small $(<\text{few}\%)$ in all the slab systems, a feature which is common to all our 1D test systems. The breakdown of the HF method suggests the slab systems are extremely sensitive to this small amount of electron correlation. In this sense, the low-density slabs are in fact systems of relatively strong correlation.

Traditional LDAs become exact in the limit of the HEG, i.e., when applied to the systems from which they were constructed. Our finite LDAs are, by definition, exact for the total energy when applied non-self-consistently to the slab systems, but it is of interest to examine the \textit{self-consistent} application of our LDAs to the slabs.

We find that in high-density slab systems the electron density is well matched due to the external potential being the dominant component in $V_{\rm{KS}}$. This becomes less so as we move to lower densities in which the `base' of the external potential becomes wider [see Fig.~\ref{Slab2e}(b) for a $2e$ slab case]. Consequently, erroneous dips and bumps form in the plateau regions of the LDA electron density.

To examine the errors in the density we analyze $V_{\rm{xc}}$. As well as missing out the long-range $V_{\rm{xc}}$ fields that are present in the exact system, we find the LDAs break down in the critical central region where the vast majority of the electron density is. We can attribute this to the exact $V_{\rm{xc}}$ being highly nonlocal in these systems whereas the LDAs only depend on the local density. 

The self-consistent energies of our slab systems are accurate with errors below $1 \%$, despite the self-consistent density being far from exact. Hence, as shown in Ref.~\onlinecite{Errors_Energy}, errors in the density can be canceled by errors inherent in the approximate energy functional. However, the derivative of the energy functional is less forgiving of these errors, leading to an inaccurate xc potential and density. 

\begin{figure}[htbp] 
\centering
\includegraphics[width=1.0\linewidth]{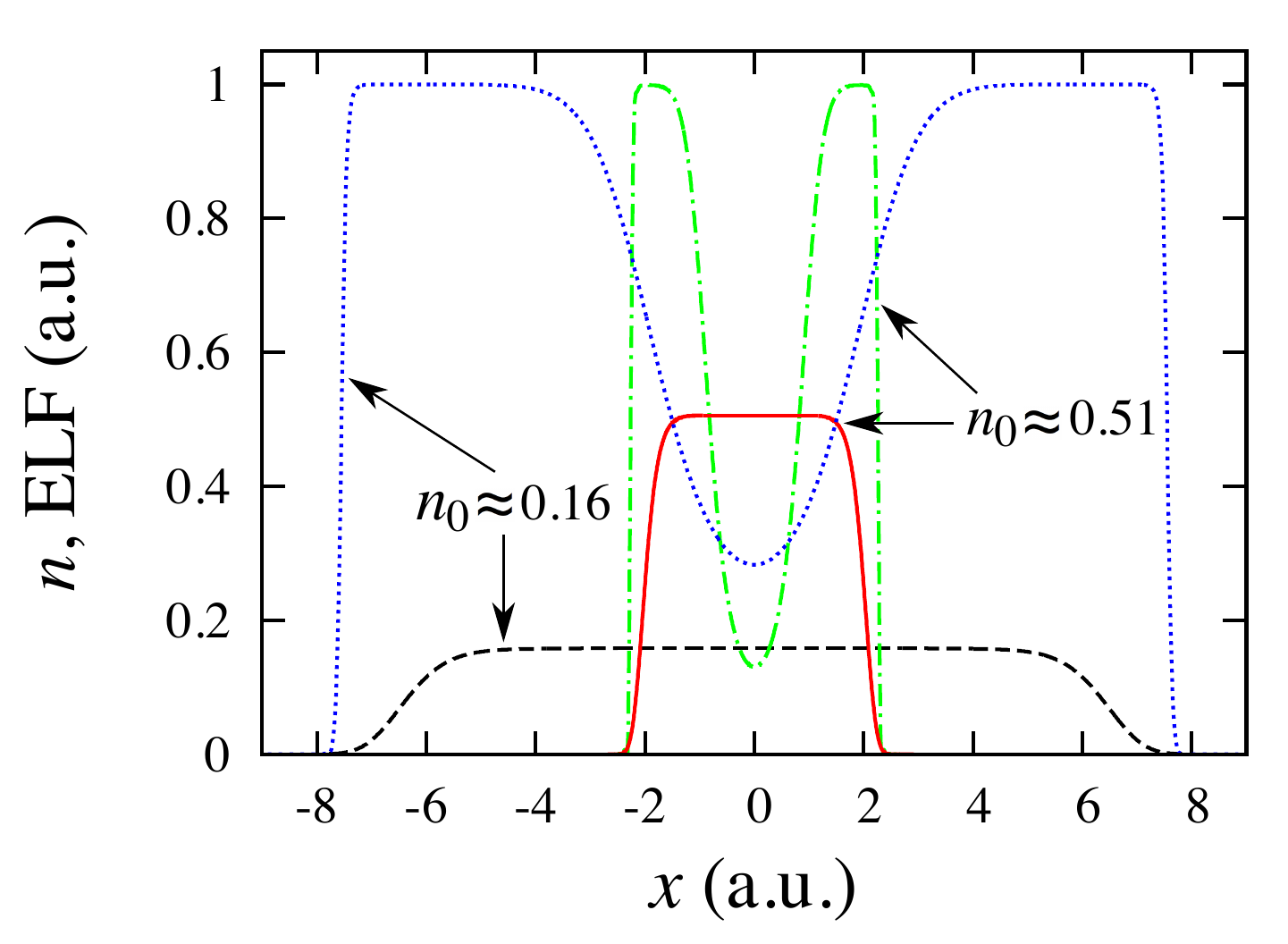}
\caption{The exact many-body electron density (dashed black line) and ELF (dotted blue line) for a low-density two-electron slab system $(n_{\rm{0}} \approx 0.16).$ Also plotted are the exact many-body electron density (solid red line) and ELF (dotted-dashed green line) for a high-density two-electron slab system $(n_{\rm{0}} \approx 0.51).$ In both systems, we find that the electrons are extremely localized towards the edges but as we approach the interface between the electrons strong delocalization occurs. In the high-density slab system, this dip in localization is deeper and occupies a greater proportion of the overall system.} 
\label{Slab_elf}
\end{figure}

Electron localization \cite{ELF2,iDEA2} is the tendency of an electron in a many-body system to exclude other electrons from its vicinity. The electron localization function (ELF) \cite{ELF,ELF2,ELF3} provides a useful indicator of localization: ELF = 1 is complete localization; i.e., the chance of finding one electron in the vicinity of another is zero. ELF ranges from 0 to 1, and a HEG has ELF = 0.5. For comparison we apply the exact ELF developed by Dobson \cite{ELF} (using our knowledge of the many-body wave function) to the $2e$ slab systems. We find that the electrons are extremely localized towards the edges of the systems but as we approach the interface between the electrons strong delocalization occurs [Fig.~\ref{Slab_elf}]. The plot shows that as we move to a high-density slab system, this dip in localization increases in depth and occupies a greater proportion of the overall system. (This is also observed in the $3e$ slab systems; however there is an extra localization peak and dip due to the third electron.)

Our results show two major differences in electron localization between the slab systems and the HEG. First, the ELF is constant across a HEG and is independent of the density. It varies between (many-electron) slab systems of different densities and is position-dependent. Second, the slab systems have regions of very high localization.
In the HEG, the ELF is defined to be 0.5 in this case, but our results (e.g. Fig.~\ref{HEG_ex}) indicate that the physical nature of the correlation (in the broad sense) in a HEG is, in fact, much more akin to that in relatively strongly localized systems -- such as our finite slab systems -- than is often supposed. That is, in a HEG, at densities much greater than those required to obtain strict localization through Wigner crystallization, a degree of localization exists which might be termed incipient Wigner crystallization.

\section{Application to exchange-dominated systems} 
In the previous section we observed the dominance of the exchange energy in the slab systems. In this section we investigate the capacity of our LDAs to describe systems dominated either by the exchange energy or by the self-interaction correction.

\subsection{Two-electron triple well} 
\begin{figure}[htbp] 
\centering
\includegraphics[width=1.0\linewidth]{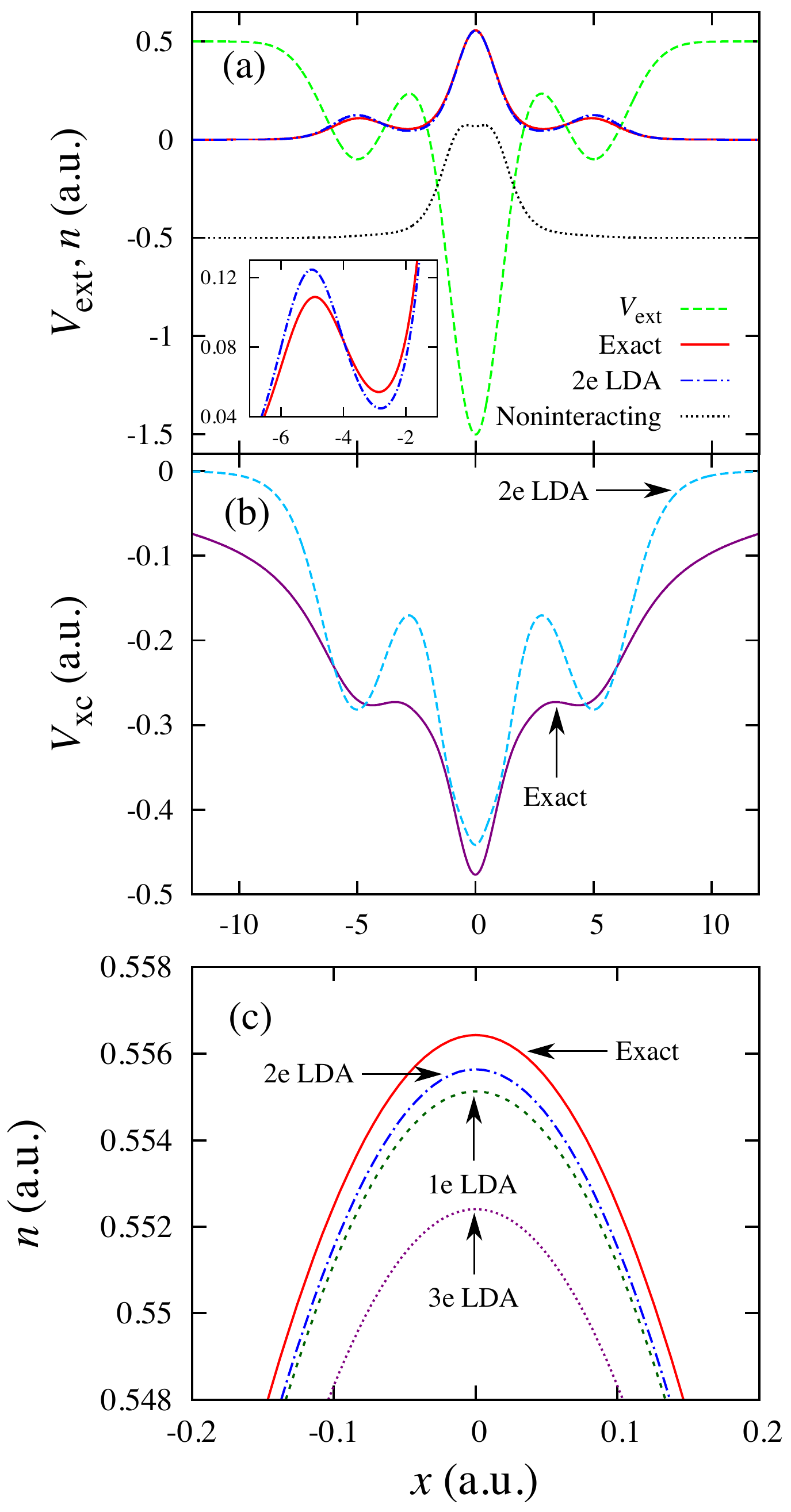}
\caption{A triple well containing two electrons. (a) A comparison of the exact many-body electron density (solid red line), the density obtained from applying the $2e$ LDA (dotted-dashed blue line), and the density obtained when we use the noninteracting approximation (dotted black line, shifted down by 0.5 to more easily distinguish between the different densities), along with the external potential (dashed green line). The LDA approximates the density remarkably well, while the noninteracting approximation incorrectly predicts both electrons occupying the central well. Inset: Close-up of the exact density and the $2e$ LDA density at the interface between the left-hand side well and the central well. (b) The exact $V_{\rm{xc}}$ (solid purple line), along with the $V_{\rm{xc}}$ obtained from applying the $2e$ LDA (dashed light-blue line). The LDA replicates the self-interaction correction remarkably well, seen in the large dips in $V_{\rm{xc}}$. However, it misses out nonlocal features present in the exact $V_{\rm{xc}}$. (c) Detail of the peak of the exact many-body electron density (solid red line) in the central well along with the densities obtained by applying the $1e$ (short-dashed dark-green line), $2e$ (dotted-dashed blue line), and $3e$ (dotted purple line) LDAs. All three LDAs accurately describe the self-interaction correction.}
\label{TripleWell}
\end{figure}

We begin the testing of the LDAs by studying a ground-state system where the electrons are \textit{highly localized}: two electrons subject to an external potential consisting of a deep, central well and two identical, shallow, side wells \footnote{See Supplemental Material for the specific parameters of our test systems.} (two-electron triple well). The exact many-body electron density, which we calculate using iDEA, is compared to the density that is obtained when we apply the $2e$ LDA self-consistently and the density obtained when we use the noninteracting approximation [Fig.~\ref{TripleWell}(a)]. The LDA does a remarkable job of matching the exact electron density. The Hartree potential acts to drive the electrons apart, with the xc potential then making the density accurate. However, the noninteracting approximation wrongly predicts both electrons occupying the central well, due to the first two single-particle energy states being lower than the potential barrier between the central well and the side wells. The HF method performs very well in this system due to strong exchange. 

\begin{table*}[t] 
\caption{\label{TripleWell_Table}The total energies and xc energies calculated self-consistently using the LDAs and their associated errors for the two-electron triple well. All three LDAs perform very well in both cases.}
\begin{ruledtabular}
\begin{tabular}{ccccccc}
  LDA & $E^{\rm{LDA}}$ (a.u.) & $\Delta E$ (a.u.) & $\% \ \text{Error}$ & $E_{\rm{xc}}^{\rm{LDA}}$ (a.u.) & $\Delta E_{\rm{xc}}$ (a.u.) & $\% \ \text{Error}$\\
  \hline
  $1e$ & -0.698 & -0.008 & -1 & -0.474 & -0.007 & -1\\
  $2e$ & -0.697 & -0.007 & -1 & -0.472 & -0.005 & -1\\
  $3e$ & -0.698 & -0.008 & -1 & -0.472 & -0.005 & -1\\
\end{tabular}
\end{ruledtabular}
\end{table*}

To understand these results we analyze the xc potential. The large dips in the exact $V_{\rm{xc}}$ \footnote{The exact $V_{\rm{xc}}$ is obtained up to an additive constant, which we choose so that $V_{\rm{xc}}$ asymptotically approaches zero as $|x| \rightarrow \infty$.} [Fig.~\ref{TripleWell}(b)] corresponding to the peaks in the electron density are primarily due to the \textit{self-interaction correction}, i.e., occurring in regions of high electron localization. The LDA does quite an extraordinary job of replicating this which explains the success in approximating the electron density. This is a particularly striking feature as traditional LDAs do not perform well in highly localized systems, as they are unable to accurately describe the self-interaction correction. The discrepancy in $V_{\rm{xc}}$ in the low density regions, at the interfaces of the wells in $V_{\rm{ext}}$, is due to the LDA being dependent on the local density and hence not accounting for nonlocal effects. These nonlocal features in the exact $V_{\rm{xc}}$ lead to, among other things, lower peaks in the density in the side wells [inset of Fig.~\ref{TripleWell}(a)]. As expected, the LDA incorrectly predicts $V_{\rm{xc}}$ decaying exponentially rather than following a Coulomb-like $-1/x$ decay.

We now look at how well each of the LDAs describe the self-interaction correction in this system. To do this we compare the electron density as predicted by each LDA to the exact many-body electron density in the highly localized central well [Fig.~\ref{TripleWell}(c)]. The $2e$ LDA is the most accurate, closely followed by the $1e$ LDA and then the $3e$ LDA. However, in general, we find that the $N$-electron LDA ($N = 1, 2, $ or $ 3$) does not necessarily perform best when applied to an $N$-electron system. In the majority of systems we study, the $1e$ LDA most accurately describes the self-interaction correction, followed by the $2e$ LDA and then the $3e$ LDA.

The final quantities we use to compare the merits of the LDAs are the approximations to $E$ and $E_{\rm{xc}}$, due to the fundamental importance of energy calculations in DFT. To do this we first calculate the exact $E$ for the two-electron triple-well system through iDEA and from this we calculate the exact $E_{\rm{xc}}$. We obtain $E = -0.690 \text{ a.u.}$ and $E_{\rm{xc}} = -0.467 \text{ a.u.}$

For each LDA we take the self-consistently calculated electron density to determine the self-consistently calculated energies. The set of self-consistently calculated $E$, $E^{\rm{LDA}}$, along with the error relative to the exact $E$, $\Delta E$, and the corresponding percentage error, $\% \ \text{error}$, are given in Table~\ref{TripleWell_Table}.  Also given are the set of self-consistently calculated $E_{\rm{xc}}$, $E_{\rm{xc}}^{\rm{LDA}}$, along with the error relative to the exact $E_{\rm{xc}}$, $\Delta E_{\rm{xc}}$, and the corresponding percentage error, $\% \ \text{error}$. The results show that all three LDAs do an impressive job of approximating $E$ and $E_{\rm{xc}}$.

\subsection{One-electron harmonic well} 
\begin{figure}[htbp] 
\centering
\includegraphics[width=1.0\linewidth]{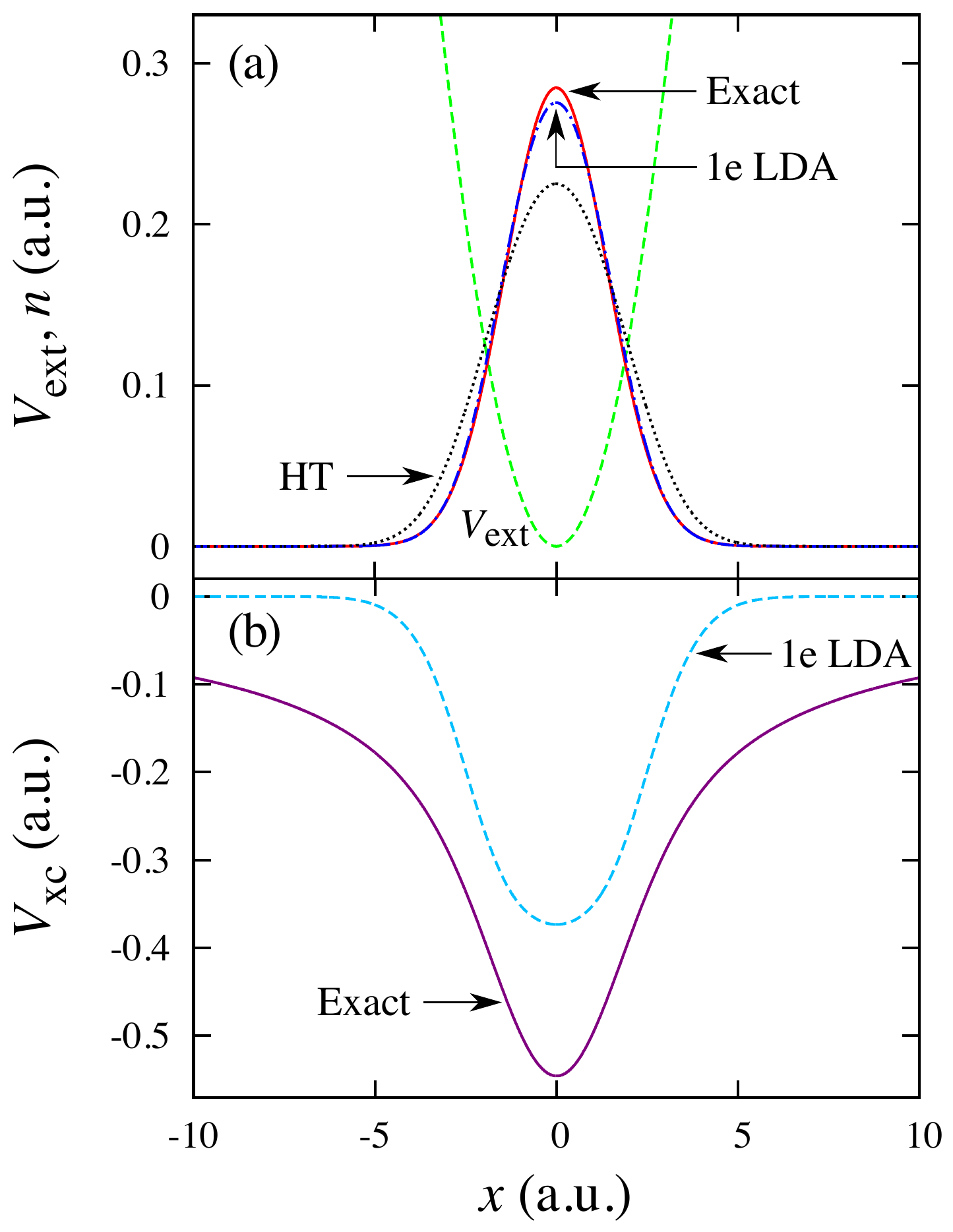}
\caption{A harmonic well containing one electron. (a) A comparison of the exact many-body electron density (solid red line), the density obtained from applying the $1e$ LDA (dotted-dashed blue line), and the density obtained when we use HT (dotted black line), along with the external potential (dashed green line). Again, the LDA approximates the density remarkably well. It captures the central peak in the density and correctly predicts its rate of decay towards the edges of the system. This is a significant improvement on HT. (b) The exact $V_{\rm{xc}}$ (solid purple line) for the one-electron harmonic well, along with the $V_{\rm{xc}}$ obtained from applying the $1e$ LDA (dashed light-blue line). $V_{\rm{xc}}$ in this system is entirely self-interaction correction and the LDA  performs well, much like it did in the two-electron triple well in which $V_{\rm{xc}}$ was mostly self-interaction correction. The LDA accurately describes the dip in $V_{\rm{xc}}$ in the center of the system; however, there is an error relative to the exact $V_{\rm{xc}}$. Again, the LDA incorrectly predicts $V_{\rm{xc}}$ decaying exponentially rather than following a Coulomb like $-1/x$ decay.}
\label{1eHW}
\end{figure}

\begin{table*}[t] 
\caption{\label{1eHW_Table}The total energies and xc energies calculated self-consistently using the LDAs and their associated errors for the one-electron harmonic well. While the $\% \ \text{Errors}$ in the total energy and the xc energy are larger than in the two-electron triple well, the relative errors are of the same order. Again, all three LDAs give similar results.}
\begin{ruledtabular}
\begin{tabular}{ccccccc}
  LDA & $E^{\rm{LDA}}$ (a.u.) & $\Delta E$ (a.u.) & $\% \ \text{Error}$ & $E_{\rm{xc}}^{\rm{LDA}}$ (a.u.) & $\Delta E_{\rm{xc}}$ (a.u.) & $\% \ \text{Error}$\\
  \hline
  $1e$ & 0.138 & 0.011 & 9 & -0.225 & 0.012 & 5\\
  $2e$ & 0.139 & 0.012 & 9 & -0.223 & 0.014 & 6\\
  $3e$ & 0.137 & 0.010 & 8 & -0.224 & 0.013 & 5\\
\end{tabular}
\end{ruledtabular}
\end{table*}

We now study a ground-state system in which exchange and correlation consist exclusively of the self-interaction correction: \textit{one} electron subject to a harmonic external potential (one-electron harmonic well). The electron behaves as a quantum harmonic oscillator with the density forming a single peak in the center of the well. This exact electron density is compared to the density that is obtained when we apply the $1e$ LDA self-consistently and the density obtained when we set $V_{\rm{xc}} = 0$, i.e., Hartree theory (HT) \footnote{The $V_{\rm{xc}} = 0$ approximation (Hartree theory) provides a benchmark against which we test the LDAs' ability to describe the self-interaction correction.} [Fig.~\ref{1eHW}(a)]. Much like in the two-electron triple well, the LDA gives a result which closely matches the exact electron density. It captures the central peak in the density and correctly predicts its rate of decay towards the edges of the system. It is worth noting that all three LDAs give very similar results with the $1e$ LDA performing the best by a small margin. We choose to only illustrate the $1e$ LDA here. HT gives a poor performance which misses out both of these features. Both the HF method and the noninteracting approximation are exact in a one-electron system.

In a one-electron system the exact $V_{\rm{xc}}$ is just the negative of the Hartree potential $V_{\rm{H}}$. Much like the LDAs' remarkable success in the two-electron triple well, in which $V_{\rm{xc}}$ is mostly self-interaction correction, it also performs well at approximating $V_{\rm{xc}}$ in this system [Fig.~\ref{1eHW}(b)]. The LDA accurately describes the dip in $V_{\rm{xc}}$ in the center of the system; however, there is an error relative to the exact $V_{\rm{xc}}$. Again, the LDA incorrectly predicts $V_{\rm{xc}}$ decaying exponentially rather than following a Coulomb like $-1/x$ decay. (We have tested the LDA in a variety of harmonic wells as we vary the angular frequency $\omega$ and we obtain similar results.)

As for the two-electron triple well, we perform energy calculations to obtain $E = 0.127 \text{ a.u.}$ and $E_{\rm{xc}} = -0.237 \text{ a.u.}$ We calculate $E^{\rm{LDA}}$ for each LDA along with $\Delta E$ and the $\% \ \text{error}$. This is displayed in Table~\ref{1eHW_Table}. Also given is the calculated $E_{\rm{xc}}^{\rm{LDA}}$ for each LDA along with $\Delta E_{\rm{xc}}$ and the $\% \ \text{error}$. While the $\% \ \text{errors}$ are noticeably larger in this system than in the two-electron triple well (see Table~\ref{TripleWell_Table}), it is the relative errors $\Delta E$ that are important. (Adding a constant to $V_{\rm{ext}}$ will change the $\% \ \text{errors}$ but not $\Delta E$.) These are of the same order as those in the two-electron triple well, with all three LDAs performing similarly. 

\subsection{Summary} 
We observe our LDA calculations to yield accurate electron densities for a variety of exchange-dominated systems, even when the LDA is constructed from one-electron systems. The most striking aspect of our LDAs are their ability to accurately describe the self-interaction correction. This is remarkable as local approximations are traditionally known to be incapable of accurately describing this feature. However, we note that some systems exhibit highly nonlocal features in the exact exchange-correlation potential, such as potential steps and other features in low density regions \cite{Steps4,Steps}. These absent nonlocal features in $V_\mathrm{xc}^\mathrm{LDA}$ can lead to inaccurate electron densities for ground-state systems, as well as for time-dependent systems; see Sec~\ref{Tunneling_system}.

\section{Application to more strongly correlated systems} 

In the previous section we observed the capacity of our LDAs to describe exchange and the self-interaction correction. We now study systems in which correlation is stronger, a feature which should challenge local approximations.  

\subsection{Two-electron harmonic wells} 

We now consider a pair of systems which demonstrate the effect on the LDAs when electron correlation increases: \textit{two} electrons confined to a harmonic external potential. First, for purposes of comparison, we consider a strongly confining harmonic external potential ($\omega = 0.4$ a.u.) so that the system is dominated by exchange, and correlation is very low (strongly confined harmonic well). We contrast this with a weakly confining harmonic external potential ($\omega = 0.01$ a.u.) in which correlation increases \textit{significantly}, as kinetic energy diminishes (weakly confined harmonic well). 

\begin{figure}[htbp]
\centering
\includegraphics[width=1.0\linewidth]{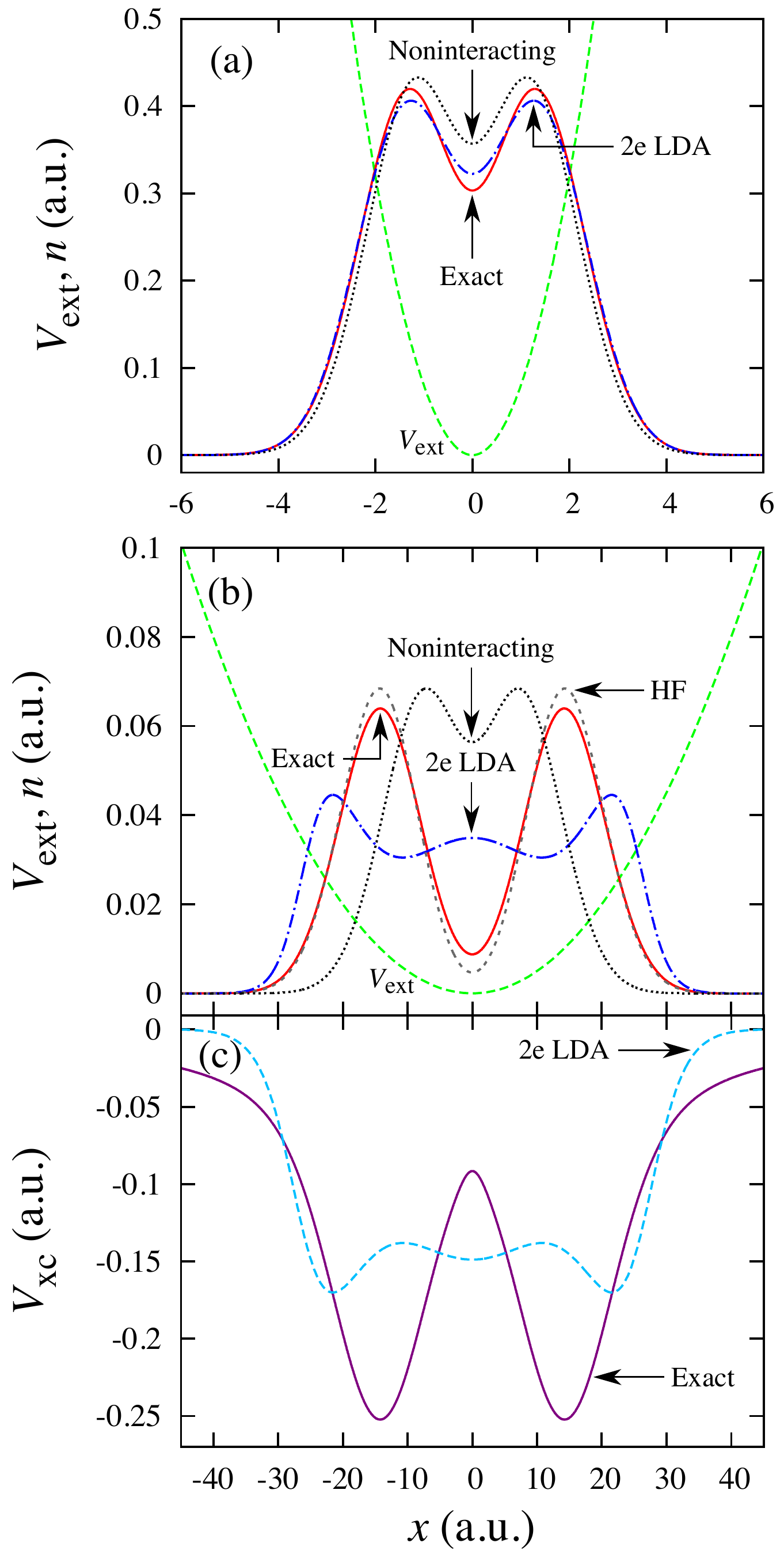}
\caption{Two-electron harmonic wells with weak and stronger correlation. (a) A comparison of the exact many-body electron density (solid red line), the density obtained from applying the $2e$ LDA (dotted-dashed blue line), and the density obtained when we use the noninteracting approximation (dotted black line), along with the external potential (dashed green line) for the strongly confined harmonic well. The LDA performs very well in this exchange-dominated system. (b) A comparison of the exact many-body electron density (solid red line), the density obtained from applying the $2e$ LDA (dotted-dashed blue line), the density obtained when we use the noninteracting approximation (dotted black line), and the density obtained when we use the HF method (short-dashed dark-gray line), along with the external potential (dashed green line) for the weakly confined harmonic well. The LDA completely breaks down in this strongly correlated system. (c) The exact $V_{\rm{xc}}$ (solid purple line) for the weakly confined harmonic well along with the $V_{\rm{xc}}$ obtained from applying the $2e$ LDA (dashed light-blue line). The LDA incorrectly predicts a central dip in $V_{\rm{xc}}$. This, along with the LDA vastly underestimating the two other dips in $V_{\rm{xc}}$, leads to the three peaks that are seen in the electron density.}
\label{2eHW}
\end{figure}

In the strongly confined harmonic well, the HF method is almost exact due to the near absence of electron correlation. The exact electron density is compared to the density that is obtained when we apply the $2e$ LDA self-consistently and the density obtained when we use the noninteracting approximation [Fig.~\ref{2eHW}(a)]. We find that the LDA performs very well in this system, which is consistent with the other two exchange-dominated systems in the previous section. (All three LDAs perform similarly.) Again, we analyze $V_{\rm{xc}}$ and find that the LDA misses out key nonlocal features, e.g., a central bump in the exact $V_{\rm{xc}}$, formed from the superposition of two steps (yielded by a single interaction term), which acts to drive the electrons further apart, leading to a discrepancy in the electron density. Even though $V_{\rm{ext}}$ is the dominant component in $V_{\rm{KS}}$, the Coulomb interaction is key to push the electrons apart, which is evident by comparing the exact density and the LDA density to the noninteracting approximation.

As we move to the weakly confined harmonic well, we find that correlation increases. This is evident in the electron density produced by the HF method becoming worse, which we compare with the exact density, the density obtained when we apply the $2e$ LDA self-consistently, and the density obtained when we use the noninteracting approximation [Fig.~\ref{2eHW}(b)]. Despite the LDA being constructed from slab systems in which correlation is significant, we find that it completely breaks down in this system. It incorrectly predicts three peaks in the electron density and appears to closely approximate a slablike system. The noninteracting approximation performs much worse than in the strongly confined harmonic.

We analyze $V_{\rm{xc}}$ and find that, unlike in the strongly confined harmonic well in which the LDA underestimated the central bump present in the exact $V_{\rm{xc}}$, it does worse in this system by incorrectly predicting a central dip in $V_{\rm{xc}}$ [Fig.~\ref{2eHW}(c)]. This, along with the LDA vastly underestimating the two other dips in $V_{\rm{xc}}$, leads to the three peaks that are seen in its approximation to the electron density. Again, the LDA incorrectly predicts an exponential decay of $V_{\rm{xc}}$ towards the system's edges.

Finally, we perform energy calculations to obtain $E = 0.068 \text{ a.u.}$ and $E_{\rm{xc}} = -0.215 \text{ a.u.}$ for the weakly confined harmonic well. We calculate $E^{\rm{LDA}}$ for each LDA along with $\Delta E$ and the $\% \ \text{error}$. This is displayed in Table~\ref{WCHW_Table}. The LDAs give good approximations to $E$ despite poor electron densities \cite{Errors_Energy}. 

We calculate $E_{\rm{xc}}^{\rm{LDA}}$ for each LDA along with $\Delta E_{\rm{xc}}$ and the $\% \ \text{error}$ for the weakly confined harmonic well. This is also displayed in Table~\ref{WCHW_Table}. Clearly the LDAs perform much worse at approximating $E_{\rm{xc}}$ than they do at approximating $E$. While we find the errors are substantially larger than in the strongly confined harmonic well (error $\sim 2\%$ for all three LDAs), one might expect a larger error on the basis of the inaccuracy of the density given by the LDAs [see Fig.~\ref{2eHW}(b) for the $2e$ LDA]. 

\begin{table*}[t] 
\caption{\label{WCHW_Table}The total energies and xc energies calculated self-consistently using the LDAs and their associated errors for the weakly confined harmonic well. The LDAs give good approximations to $E$ despite poor electron densities, which we attribute to a cancellation of errors. They perform much worse at approximating $E_{\rm{xc}}$.}
\begin{ruledtabular}
\begin{tabular}{ccccccc}
  LDA & $E^{\rm{LDA}}$ (a.u.) & $\Delta E$ (a.u.) & $\% \ \text{Error}$ & $E_{\rm{xc}}^{\rm{LDA}}$ (a.u.) & $\Delta E_{\rm{xc}}$ (a.u.) & $\% \ \text{Error}$\\
  \hline
  $1e$ & 0.072 & 0.004 & 6 & -0.182 & 0.033 & 15\\
  $2e$ & 0.066 & -0.002 & -3 & -0.186 & 0.029 & 13\\
  $3e$ & 0.063 & -0.005 & -7 & -0.191 & 0.024 & 11\\
\end{tabular}
\end{ruledtabular}
\end{table*}

\subsection{Tunneling system} \label{Tunneling_system}
\begin{figure}[H] 
\centering
\includegraphics[width=1.0\linewidth]{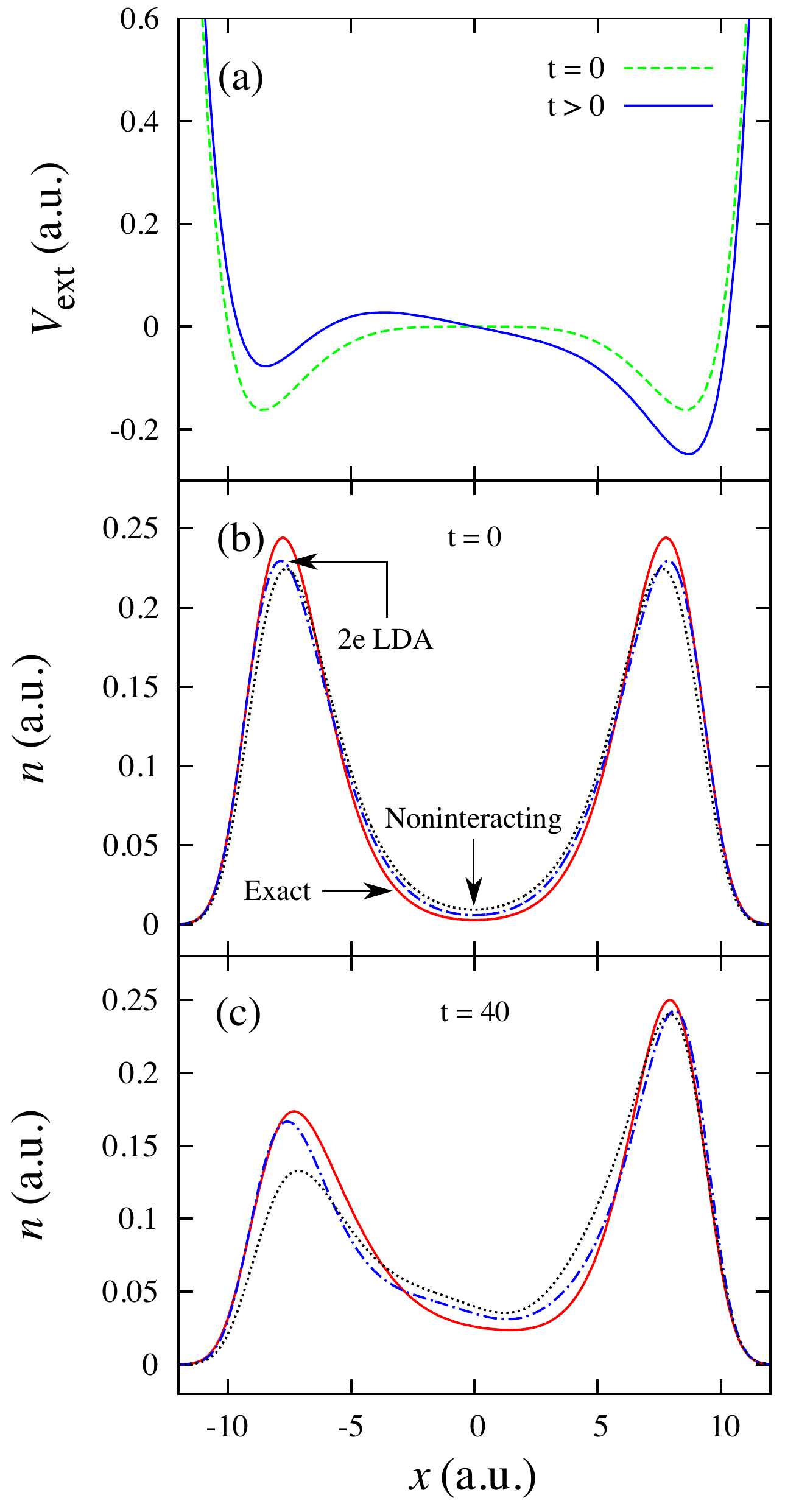}
\caption{A tunneling system containing two electrons. (a) The unperturbed external potential $V_{\rm{ext}} = \alpha x^{10} - \beta x^{4}$, $t = 0$ (dashed green line), and the perturbed external potential with the electric field $- \varepsilon x$, applied for $t > 0$ (solid blue line). (b) A comparison of the exact many-body electron density (solid red line), the density obtained from applying the $2e$ LDA (dotted-dashed blue line), and the density obtained when we use the noninteracting approximation (dotted black line) for the system's ground-state, $t = 0$. The Pauli exclusion principle, combined with the Coulomb repulsion, forces the electrons to localize in opposite wells resulting in a small-density barrier (central) region. The LDA and the noninteracting approximation both match this well. (c) A comparison of the exact many-body electron density (solid red line), the density obtained from applying the $2e$ LDA (dotted-dashed blue line), and the density obtained when we use the noninteracting approximation (dotted black line) at a later time, $t = 40$ a.u., once there has been sufficient tunneling. While the LDA still manages to replicate the exact density well, it fails in the critical central region which indicates that the tunneling rate is too high.}
\label{Tunnelling}
\end{figure}

We now extend our study to a \textit{highly correlated} time-dependent system in which there is \textit{strong current flow}: two electrons confined to an external potential consisting of two wells separated by a long flat barrier, $V_{\rm{ext}} = \alpha x^{10} - \beta x^{4}$, where $\alpha = 5 $ x $ 10^{-11}$ a.u.\ and $\beta = 0.5 $ x $ 10^{-4}$ a.u. For $t > 0$ a perturbing electric field $V_{\rm{pert}} = \varepsilon x$, where $\varepsilon = -0.01$, is applied [Fig.~\ref{Tunnelling}(a)] to induce quantum tunneling (tunneling system)\cite{iDEA}.

The Pauli exclusion principle, combined with the Coulomb repulsion, forces the electrons to localize in opposite wells resulting in a small-density barrier (central) region. This is well matched, both when we apply the $2e$ LDA and when we use the noninteracting approximation [Fig.~\ref{Tunnelling}(b)]. We apply the HF method and find this to be an exchange-dominated system. Again, the LDA accurately describes the large self-interaction correction present in the highly localized wells. As in the strongly confined harmonic well, there is a central bump present in the exact $V_{\rm{xc}}$, which is due to the superposition of two steps. The LDA misses out this key feature, which acts to drive the electrons apart, leading to higher peaks in the exact electron density.

The application of the electric field initially causes the electrons to oscillate within their respective wells. Eventually the electron in the left hand well begins to tunnel through the potential barrier towards the right hand well. Correlation increases as the electrons begin to explore different orbitals. We apply the LDA adiabatically, $V_{\rm{xc}}^{\rm{ALDA}}[n](x,t) = V_{\rm{xc}}^{\rm{LDA}}[n(t)](x)$, to examine how well it approximates the dynamic electron density once there has been sufficient tunneling $(t = 40$ a.u.$)$, along with the result that is obtained when we use the noninteracting approximation [Fig.~\ref{Tunnelling}(c)]. While the LDA still manages to replicate the exact density well, it fails in the critical central region which indicates that the tunneling rate is too high. However, it is an improvement on the density that is obtained when we neglect the Coulomb interaction.

\begin{figure}[htbp] 
\centering
\includegraphics[width=1.0\linewidth]{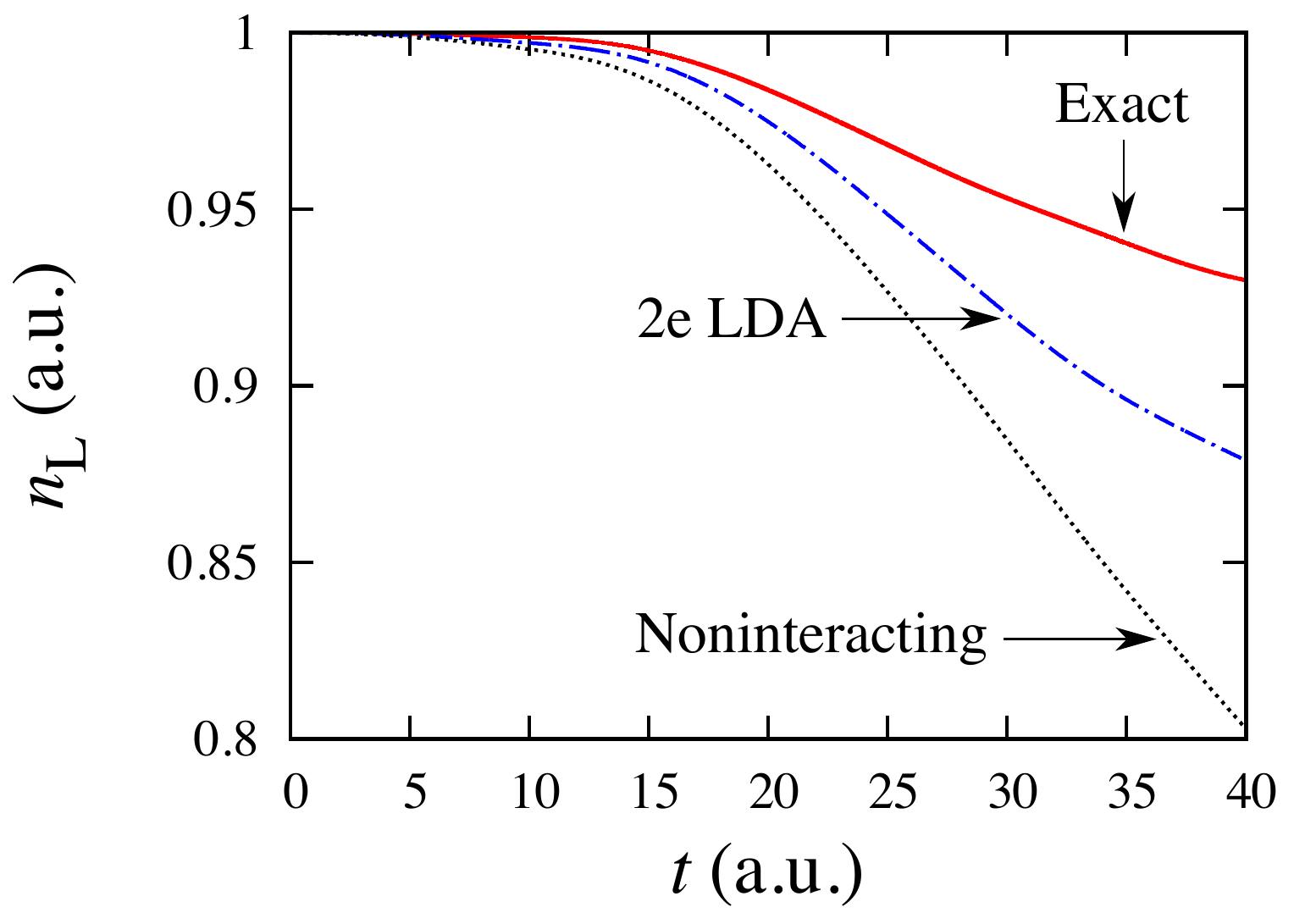}
\caption{The exact total electron number on the LHS ($x<0$) of the system $n_{\rm{L}}$ (solid red line), the approximation produced from applying the $2e$ LDA adiabatically (dotted-dashed blue line), and the one obtained when we use the noninteracting approximation (dotted black line). The tunneling rates are therefore given by the gradients of the curves. It is clear that the LDA overestimates the rate of tunneling. By taking the gradients of the three curves, we measure the magnitude of the LDA tunneling rate to be, on average, nearly twice that of the exact tunneling rate. At early times, this error is due to the LDA missing out key ground-state features. At later times, it is primarily due to increasing correlation as the electrons explore different orbitals. The noninteracting approximation further overestimates the tunneling rate.}
\label{TunnellingRate}
\end{figure}

To explore this we first define the tunneling rate as the rate at which the total electron density on the left hand side (LHS, $x<0$) of the system decreases with time. (This is deemed to be a sufficient approximation as the electrons start in a highly localized ground state.) We now plot the exact total electron density in the LHS as a function of time, the approximation produced from applying the LDA, and the result obtained when we use the noninteracting approximation [Fig.~\ref{TunnellingRate}]. In all three cases the tunneling rate increases as the LHS electron gains kinetic energy, before decreasing in response to an increase in the Coulomb repulsion. It is clear that the LDA overpredicts the rate of tunneling. By taking the gradients of the three curves, we measure the magnitude of the LDA tunneling rate to be, on average, nearly twice that of the exact tunneling, although this is a large reduction in the erroneous tunneling rate obtained when we use the noninteracting approximation.

Dynamic potential steps have previously been shown to be important nonlocal features which give rise to accurate electron densities\cite{Steps,Steps2,Steps3}. We observe a dynamic step to grow in the exact $V_{\rm{xc}}$ (and hence $V_{\rm{KS}}$) in the central density minimum, which in turn controls the tunneling rate. Unsurprisingly, this characteristic is missing from the LDA $V_{\rm{xc}}$. In order to slow the tunneling rate to an appropriate amount, a better approximate functional will be needed; one that takes into account the current density, which is particularly sensitive to interaction in this system. We observe this through the LDA current density quickly deviating from the exact current density, which is reflected in the time-dependent density. 

We find that at early times, errors in the time-dependent density depend heavily on how well the ground-state is approximated. Therefore, we find that accurately describing ground-state features is crucial. At later times, the error in the LDA density grows primarily due to increasing correlation.

\subsection{Summary} 
Similar to traditional local approximations, we have found that our LDAs are unable to accurately describe systems in which correlation is significant. The transition from the strongly confined harmonic well to the weakly confined harmonic well demonstrates that while the LDAs can successfully be applied to exchange-dominated systems, an increase in the correlation energy causes them to become severely inaccurate. This is also observed in the tunneling system, in which starting from a highly localized ground-state, the approximation to the electron density becomes worse as correlation increases with time. Therefore, despite the low-density slab systems being strongly correlated, correlation effects in test systems do not appear to be captured by the LDAs.

\section{Conclusions} 
We have introduced a set of three LDAs constructed from the exact properties of finite systems consisting of as few as one electron, as an alternative to the homogeneous electron gas. The three LDAs are remarkably similar to one another. By analyzing calculations for a HEG using a closely related 1D interaction \cite{LDA_QMC}, we conclude that our three LDAs are also similar to a HEG-based LDA, contradicting the common idea that localization differs greatly in a HEG from that in finite systems. Extending to 3D, we find that an LDA constructed from finite systems containing just one electron is feasible.

One of the most surprising features of our 1D LDAs is their ability to accurately describe the self-interaction correction. When combined with the Hartree potential, this leads to good electron densities, especially in regions of high electron localization. Much like 3D LDAs, we find that our LDAs perform well in test systems dominated by the exchange energy, but are much less reliable when correlation is stronger. However, by definition, the LDAs omit nonlocal features in the xc functional, such as steps, which are needed in certain systems to give accurate electron densities. 

\begin{acknowledgments}
We acknowledge funding from the Ogden Trust, Engineering  and Physical Sciences Research Council (EPSRC), and the York Centre for Quantum Technologies (YCQT). We thank Phil Hasnip and Matt Probert for helpful discussions.
\end{acknowledgments}

\bibliography{Entwistle_2016}

\begin{thebibliography}{53}%
\makeatletter
\providecommand \@ifxundefined [1]{%
 \@ifx{#1\undefined}
}%
\providecommand \@ifnum [1]{%
 \ifnum #1\expandafter \@firstoftwo
 \else \expandafter \@secondoftwo
 \fi
}%
\providecommand \@ifx [1]{%
 \ifx #1\expandafter \@firstoftwo
 \else \expandafter \@secondoftwo
 \fi
}%
\providecommand \natexlab [1]{#1}%
\providecommand \enquote  [1]{``#1''}%
\providecommand \bibnamefont  [1]{#1}%
\providecommand \bibfnamefont [1]{#1}%
\providecommand \citenamefont [1]{#1}%
\providecommand \href@noop [0]{\@secondoftwo}%
\providecommand \href [0]{\begingroup \@sanitize@url \@href}%
\providecommand \@href[1]{\@@startlink{#1}\@@href}%
\providecommand \@@href[1]{\endgroup#1\@@endlink}%
\providecommand \@sanitize@url [0]{\catcode `\\12\catcode `\$12\catcode
  `\&12\catcode `\#12\catcode `\^12\catcode `\_12\catcode `\%12\relax}%
\providecommand \@@startlink[1]{}%
\providecommand \@@endlink[0]{}%
\providecommand \url  [0]{\begingroup\@sanitize@url \@url }%
\providecommand \@url [1]{\endgroup\@href {#1}{\urlprefix }}%
\providecommand \urlprefix  [0]{URL }%
\providecommand \Eprint [0]{\href }%
\providecommand \doibase [0]{http://dx.doi.org/}%
\providecommand \selectlanguage [0]{\@gobble}%
\providecommand \bibinfo  [0]{\@secondoftwo}%
\providecommand \bibfield  [0]{\@secondoftwo}%
\providecommand \translation [1]{[#1]}%
\providecommand \BibitemOpen [0]{}%
\providecommand \bibitemStop [0]{}%
\providecommand \bibitemNoStop [0]{.\EOS\space}%
\providecommand \EOS [0]{\spacefactor3000\relax}%
\providecommand \BibitemShut  [1]{\csname bibitem#1\endcsname}%
\let\auto@bib@innerbib\@empty
%</preamble>
\bibitem [{\citenamefont {Hohenberg}\ and\ \citenamefont {Kohn}(1964)}]{DFT}%
  \BibitemOpen
  \bibfield  {author} {\bibinfo {author} {\bibfnamefont {P.}~\bibnamefont
  {Hohenberg}}\ and\ \bibinfo {author} {\bibfnamefont {W.}~\bibnamefont
  {Kohn}},\ }\href {\doibase 10.1103/PhysRev.136.B864} {\bibfield  {journal}
  {\bibinfo  {journal} {Phys. Rev.}\ }\textbf {\bibinfo {volume} {136}},\
  \bibinfo {pages} {B864} (\bibinfo {year} {1964})}\BibitemShut {NoStop}%
\bibitem [{\citenamefont {Kohn}\ and\ \citenamefont {Sham}(1965)}]{KS_LDA}%
  \BibitemOpen
  \bibfield  {author} {\bibinfo {author} {\bibfnamefont {W.}~\bibnamefont
  {Kohn}}\ and\ \bibinfo {author} {\bibfnamefont {L.~J.}\ \bibnamefont
  {Sham}},\ }\href {\doibase 10.1103/PhysRev.140.A1133} {\bibfield  {journal}
  {\bibinfo  {journal} {Phys. Rev.}\ }\textbf {\bibinfo {volume} {140}},\
  \bibinfo {pages} {A1133} (\bibinfo {year} {1965})}\BibitemShut {NoStop}%
\bibitem [{\citenamefont {Ceperley}\ and\ \citenamefont
  {Alder}(1980)}]{QMC_HEG}%
  \BibitemOpen
  \bibfield  {author} {\bibinfo {author} {\bibfnamefont {D.~M.}\ \bibnamefont
  {Ceperley}}\ and\ \bibinfo {author} {\bibfnamefont {B.~J.}\ \bibnamefont
  {Alder}},\ }\href {\doibase 10.1103/PhysRevLett.45.566} {\bibfield  {journal}
  {\bibinfo  {journal} {Phys. Rev. Lett.}\ }\textbf {\bibinfo {volume} {45}},\
  \bibinfo {pages} {566} (\bibinfo {year} {1980})}\BibitemShut {NoStop}%
\bibitem [{\citenamefont {Dreizler}\ and\ \citenamefont
  {Gross}(2012)}]{LDA_Success}%
  \BibitemOpen
  \bibfield  {author} {\bibinfo {author} {\bibfnamefont {R.}~\bibnamefont
  {Dreizler}}\ and\ \bibinfo {author} {\bibfnamefont {E.}~\bibnamefont
  {Gross}},\ }\href {https://books.google.co.uk/books?id=t6PvCAAAQBAJ} {\emph
  {\bibinfo {title} {Density Functional Theory: An Approach to the Quantum
  Many-Body Problem}}}\ (\bibinfo  {publisher} {Springer Berlin Heidelberg},\
  \bibinfo {year} {2012})\BibitemShut {NoStop}%
\bibitem [{\citenamefont {Parr}\ and\ \citenamefont
  {Yang}(1994)}]{LDA_Success2}%
  \BibitemOpen
  \bibfield  {author} {\bibinfo {author} {\bibfnamefont {R.}~\bibnamefont
  {Parr}}\ and\ \bibinfo {author} {\bibfnamefont {W.}~\bibnamefont {Yang}},\
  }\href {https://books.google.co.uk/books?id=mxiOngEACAAJ} {\emph {\bibinfo
  {title} {Density-Functional Theory of Atoms and Molecules}}},\ International
  Series of Monographs on Chemistry\ (\bibinfo  {publisher} {Oxford University
  Press, Oxford},\ \bibinfo {year} {1994})\BibitemShut {NoStop}%
\bibitem [{\citenamefont {Perdew}\ and\ \citenamefont
  {Zunger}(1981)}]{PerdewZunger}%
  \BibitemOpen
  \bibfield  {author} {\bibinfo {author} {\bibfnamefont {J.~P.}\ \bibnamefont
  {Perdew}}\ and\ \bibinfo {author} {\bibfnamefont {A.}~\bibnamefont
  {Zunger}},\ }\href {\doibase 10.1103/PhysRevB.23.5048} {\bibfield  {journal}
  {\bibinfo  {journal} {Phys. Rev. B}\ }\textbf {\bibinfo {volume} {23}},\
  \bibinfo {pages} {5048} (\bibinfo {year} {1981})}\BibitemShut {NoStop}%
\bibitem [{\citenamefont {Tozer}\ and\ \citenamefont {Handy}(1998)}]{SIC}%
  \BibitemOpen
  \bibfield  {author} {\bibinfo {author} {\bibfnamefont {D.~J.}\ \bibnamefont
  {Tozer}}\ and\ \bibinfo {author} {\bibfnamefont {N.~C.}\ \bibnamefont
  {Handy}},\ }\href@noop {} {\bibfield  {journal} {\bibinfo  {journal} {J.
  Chem. Phys}\ }\textbf {\bibinfo {volume} {109}},\ \bibinfo {pages} {10180}
  (\bibinfo {year} {1998})}\BibitemShut {NoStop}%
\bibitem [{\citenamefont {Dreuw}\ \emph {et~al.}(2003)\citenamefont {Dreuw},
  \citenamefont {Weisman},\ and\ \citenamefont {Head-Gordon}}]{SIC2}%
  \BibitemOpen
  \bibfield  {author} {\bibinfo {author} {\bibfnamefont {A.}~\bibnamefont
  {Dreuw}}, \bibinfo {author} {\bibfnamefont {J.~L.}\ \bibnamefont {Weisman}},
  \ and\ \bibinfo {author} {\bibfnamefont {M.}~\bibnamefont {Head-Gordon}},\
  }\href@noop {} {\bibfield  {journal} {\bibinfo  {journal} {J. Chem. Phys}\
  }\textbf {\bibinfo {volume} {119}},\ \bibinfo {pages} {2943} (\bibinfo {year}
  {2003})}\BibitemShut {NoStop}%
\bibitem [{\citenamefont {Almbladh}\ and\ \citenamefont {von
  Barth}(1985)}]{CoulombDecay/ExpDecay}%
  \BibitemOpen
  \bibfield  {author} {\bibinfo {author} {\bibfnamefont {C.-O.}\ \bibnamefont
  {Almbladh}}\ and\ \bibinfo {author} {\bibfnamefont {U.}~\bibnamefont {von
  Barth}},\ }\href {\doibase 10.1103/PhysRevB.31.3231} {\bibfield  {journal}
  {\bibinfo  {journal} {Phys. Rev. B}\ }\textbf {\bibinfo {volume} {31}},\
  \bibinfo {pages} {3231} (\bibinfo {year} {1985})}\BibitemShut {NoStop}%
\bibitem [{\citenamefont {Levy}\ \emph {et~al.}(1984)\citenamefont {Levy},
  \citenamefont {Perdew},\ and\ \citenamefont {Sahni}}]{CoulombDecay}%
  \BibitemOpen
  \bibfield  {author} {\bibinfo {author} {\bibfnamefont {M.}~\bibnamefont
  {Levy}}, \bibinfo {author} {\bibfnamefont {J.~P.}\ \bibnamefont {Perdew}}, \
  and\ \bibinfo {author} {\bibfnamefont {V.}~\bibnamefont {Sahni}},\ }\href
  {\doibase 10.1103/PhysRevA.30.2745} {\bibfield  {journal} {\bibinfo
  {journal} {Phys. Rev. A}\ }\textbf {\bibinfo {volume} {30}},\ \bibinfo
  {pages} {2745} (\bibinfo {year} {1984})}\BibitemShut {NoStop}%
\bibitem [{\citenamefont {Jones}\ and\ \citenamefont
  {Gunnarsson}(1985)}]{Errors_Orbitals}%
  \BibitemOpen
  \bibfield  {author} {\bibinfo {author} {\bibfnamefont {R.~O.}\ \bibnamefont
  {Jones}}\ and\ \bibinfo {author} {\bibfnamefont {O.}~\bibnamefont
  {Gunnarsson}},\ }\href {\doibase 10.1103/PhysRevLett.55.107} {\bibfield
  {journal} {\bibinfo  {journal} {Phys. Rev. Lett.}\ }\textbf {\bibinfo
  {volume} {55}},\ \bibinfo {pages} {107} (\bibinfo {year} {1985})}\BibitemShut
  {NoStop}%
\bibitem [{\citenamefont {Runge}\ and\ \citenamefont {Gross}(1984)}]{TDDFT}%
  \BibitemOpen
  \bibfield  {author} {\bibinfo {author} {\bibfnamefont {E.}~\bibnamefont
  {Runge}}\ and\ \bibinfo {author} {\bibfnamefont {E.~K.~U.}\ \bibnamefont
  {Gross}},\ }\href {\doibase 10.1103/PhysRevLett.52.997} {\bibfield  {journal}
  {\bibinfo  {journal} {Phys. Rev. Lett.}\ }\textbf {\bibinfo {volume} {52}},\
  \bibinfo {pages} {997} (\bibinfo {year} {1984})}\BibitemShut {NoStop}%
\bibitem [{\citenamefont {van Leeuwen}(1999)}]{TDDFT2}%
  \BibitemOpen
  \bibfield  {author} {\bibinfo {author} {\bibfnamefont {R.}~\bibnamefont {van
  Leeuwen}},\ }\href {\doibase 10.1103/PhysRevLett.82.3863} {\bibfield
  {journal} {\bibinfo  {journal} {Phys. Rev. Lett.}\ }\textbf {\bibinfo
  {volume} {82}},\ \bibinfo {pages} {3863} (\bibinfo {year}
  {1999})}\BibitemShut {NoStop}%
\bibitem [{\citenamefont {Onida}\ \emph {et~al.}(2002)\citenamefont {Onida},
  \citenamefont {Reining},\ and\ \citenamefont {Rubio}}]{ALDA}%
  \BibitemOpen
  \bibfield  {author} {\bibinfo {author} {\bibfnamefont {G.}~\bibnamefont
  {Onida}}, \bibinfo {author} {\bibfnamefont {L.}~\bibnamefont {Reining}}, \
  and\ \bibinfo {author} {\bibfnamefont {A.}~\bibnamefont {Rubio}},\ }\href
  {\doibase 10.1103/RevModPhys.74.601} {\bibfield  {journal} {\bibinfo
  {journal} {Rev. Mod. Phys.}\ }\textbf {\bibinfo {volume} {74}},\ \bibinfo
  {pages} {601} (\bibinfo {year} {2002})}\BibitemShut {NoStop}%
\bibitem [{\citenamefont {Lein}\ \emph {et~al.}(2000)\citenamefont {Lein},
  \citenamefont {Gross},\ and\ \citenamefont {Perdew}}]{ALDA2}%
  \BibitemOpen
  \bibfield  {author} {\bibinfo {author} {\bibfnamefont {M.}~\bibnamefont
  {Lein}}, \bibinfo {author} {\bibfnamefont {E.~K.~U.}\ \bibnamefont {Gross}},
  \ and\ \bibinfo {author} {\bibfnamefont {J.~P.}\ \bibnamefont {Perdew}},\
  }\href {\doibase 10.1103/PhysRevB.61.13431} {\bibfield  {journal} {\bibinfo
  {journal} {Phys. Rev. B}\ }\textbf {\bibinfo {volume} {61}},\ \bibinfo
  {pages} {13431} (\bibinfo {year} {2000})}\BibitemShut {NoStop}%
\bibitem [{\citenamefont {Gritsenko}\ \emph {et~al.}(2000)\citenamefont
  {Gritsenko}, \citenamefont {Van~Gisbergen}, \citenamefont {G{\"o}rling},
  \citenamefont {Baerends} \emph {et~al.}}]{ALDA3}%
  \BibitemOpen
  \bibfield  {author} {\bibinfo {author} {\bibfnamefont {O.}~\bibnamefont
  {Gritsenko}}, \bibinfo {author} {\bibfnamefont {S.}~\bibnamefont
  {Van~Gisbergen}}, \bibinfo {author} {\bibfnamefont {A.}~\bibnamefont
  {G{\"o}rling}}, \bibinfo {author} {\bibfnamefont {E.}~\bibnamefont
  {Baerends}},  \emph {et~al.},\ }\href@noop {} {\bibfield  {journal} {\bibinfo
   {journal} {J. Chem. Phys}\ }\textbf {\bibinfo {volume} {113}},\ \bibinfo
  {pages} {8478} (\bibinfo {year} {2000})}\BibitemShut {NoStop}%
\bibitem [{\citenamefont {Maitra}\ \emph {et~al.}(2002)\citenamefont {Maitra},
  \citenamefont {Burke},\ and\ \citenamefont {Woodward}}]{ALDA4}%
  \BibitemOpen
  \bibfield  {author} {\bibinfo {author} {\bibfnamefont {N.~T.}\ \bibnamefont
  {Maitra}}, \bibinfo {author} {\bibfnamefont {K.}~\bibnamefont {Burke}}, \
  and\ \bibinfo {author} {\bibfnamefont {C.}~\bibnamefont {Woodward}},\ }\href
  {\doibase 10.1103/PhysRevLett.89.023002} {\bibfield  {journal} {\bibinfo
  {journal} {Phys. Rev. Lett.}\ }\textbf {\bibinfo {volume} {89}},\ \bibinfo
  {pages} {023002} (\bibinfo {year} {2002})}\BibitemShut {NoStop}%
\bibitem [{\citenamefont {Burke}\ \emph {et~al.}(2005)\citenamefont {Burke},
  \citenamefont {Werschnik},\ and\ \citenamefont {Gross}}]{ALDA5}%
  \BibitemOpen
  \bibfield  {author} {\bibinfo {author} {\bibfnamefont {K.}~\bibnamefont
  {Burke}}, \bibinfo {author} {\bibfnamefont {J.}~\bibnamefont {Werschnik}}, \
  and\ \bibinfo {author} {\bibfnamefont {E.}~\bibnamefont {Gross}},\
  }\href@noop {} {\bibfield  {journal} {\bibinfo  {journal} {J. Chem. Phys}\
  }\textbf {\bibinfo {volume} {123}},\ \bibinfo {pages} {062206} (\bibinfo
  {year} {2005})}\BibitemShut {NoStop}%
\bibitem [{\citenamefont {Varsano}\ \emph {et~al.}(2008)\citenamefont
  {Varsano}, \citenamefont {Marini},\ and\ \citenamefont {Rubio}}]{ALDA6}%
  \BibitemOpen
  \bibfield  {author} {\bibinfo {author} {\bibfnamefont {D.}~\bibnamefont
  {Varsano}}, \bibinfo {author} {\bibfnamefont {A.}~\bibnamefont {Marini}}, \
  and\ \bibinfo {author} {\bibfnamefont {A.}~\bibnamefont {Rubio}},\ }\href
  {\doibase 10.1103/PhysRevLett.101.133002} {\bibfield  {journal} {\bibinfo
  {journal} {Phys. Rev. Lett.}\ }\textbf {\bibinfo {volume} {101}},\ \bibinfo
  {pages} {133002} (\bibinfo {year} {2008})}\BibitemShut {NoStop}%
\bibitem [{\citenamefont {Jones}\ and\ \citenamefont
  {Gunnarsson}(1989)}]{ALDA7}%
  \BibitemOpen
  \bibfield  {author} {\bibinfo {author} {\bibfnamefont {R.~O.}\ \bibnamefont
  {Jones}}\ and\ \bibinfo {author} {\bibfnamefont {O.}~\bibnamefont
  {Gunnarsson}},\ }\href {\doibase 10.1103/RevModPhys.61.689} {\bibfield
  {journal} {\bibinfo  {journal} {Rev. Mod. Phys.}\ }\textbf {\bibinfo {volume}
  {61}},\ \bibinfo {pages} {689} (\bibinfo {year} {1989})}\BibitemShut
  {NoStop}%
\bibitem [{\citenamefont {Pickett}(1989)}]{ALDA8}%
  \BibitemOpen
  \bibfield  {author} {\bibinfo {author} {\bibfnamefont {W.~E.}\ \bibnamefont
  {Pickett}},\ }\href {\doibase http://dx.doi.org/10.1016/0167-7977(89)90002-6}
  {\bibfield  {journal} {\bibinfo  {journal} {Computer Physics Reports}\
  }\textbf {\bibinfo {volume} {9}},\ \bibinfo {pages} {115 } (\bibinfo {year}
  {1989})}\BibitemShut {NoStop}%
\bibitem [{\citenamefont {Gonze}\ \emph {et~al.}(1997)\citenamefont {Gonze},
  \citenamefont {Ghosez},\ and\ \citenamefont {Godby}}]{ALDA9}%
  \BibitemOpen
  \bibfield  {author} {\bibinfo {author} {\bibfnamefont {X.}~\bibnamefont
  {Gonze}}, \bibinfo {author} {\bibfnamefont {P.}~\bibnamefont {Ghosez}}, \
  and\ \bibinfo {author} {\bibfnamefont {R.~W.}\ \bibnamefont {Godby}},\ }\href
  {\doibase 10.1103/PhysRevLett.78.294} {\bibfield  {journal} {\bibinfo
  {journal} {Phys. Rev. Lett.}\ }\textbf {\bibinfo {volume} {78}},\ \bibinfo
  {pages} {294} (\bibinfo {year} {1997})}\BibitemShut {NoStop}%
\bibitem [{\citenamefont {Di~Ventra}\ \emph {et~al.}(2000)\citenamefont
  {Di~Ventra}, \citenamefont {Pantelides},\ and\ \citenamefont
  {Lang}}]{QuantumTransport1}%
  \BibitemOpen
  \bibfield  {author} {\bibinfo {author} {\bibfnamefont {M.}~\bibnamefont
  {Di~Ventra}}, \bibinfo {author} {\bibfnamefont {S.~T.}\ \bibnamefont
  {Pantelides}}, \ and\ \bibinfo {author} {\bibfnamefont {N.~D.}\ \bibnamefont
  {Lang}},\ }\href {\doibase 10.1103/PhysRevLett.84.979} {\bibfield  {journal}
  {\bibinfo  {journal} {Phys. Rev. Lett.}\ }\textbf {\bibinfo {volume} {84}},\
  \bibinfo {pages} {979} (\bibinfo {year} {2000})}\BibitemShut {NoStop}%
\bibitem [{\citenamefont {Koentopp}\ \emph {et~al.}(2008)\citenamefont
  {Koentopp}, \citenamefont {Chang}, \citenamefont {Burke},\ and\ \citenamefont
  {Car}}]{QuantumTransport2}%
  \BibitemOpen
  \bibfield  {author} {\bibinfo {author} {\bibfnamefont {M.}~\bibnamefont
  {Koentopp}}, \bibinfo {author} {\bibfnamefont {C.}~\bibnamefont {Chang}},
  \bibinfo {author} {\bibfnamefont {K.}~\bibnamefont {Burke}}, \ and\ \bibinfo
  {author} {\bibfnamefont {R.}~\bibnamefont {Car}},\ }\href
  {http://stacks.iop.org/0953-8984/20/i=8/a=083203} {\bibfield  {journal}
  {\bibinfo  {journal} {Journal of Physics: Condensed Matter}\ }\textbf
  {\bibinfo {volume} {20}},\ \bibinfo {pages} {083203} (\bibinfo {year}
  {2008})}\BibitemShut {NoStop}%
\bibitem [{\citenamefont {Helbig}\ \emph {et~al.}(2011)\citenamefont {Helbig},
  \citenamefont {Fuks}, \citenamefont {Casula}, \citenamefont {Verstraete},
  \citenamefont {Marques}, \citenamefont {Tokatly},\ and\ \citenamefont
  {Rubio}}]{LDA_QMC}%
  \BibitemOpen
  \bibfield  {author} {\bibinfo {author} {\bibfnamefont {N.}~\bibnamefont
  {Helbig}}, \bibinfo {author} {\bibfnamefont {J.~I.}\ \bibnamefont {Fuks}},
  \bibinfo {author} {\bibfnamefont {M.}~\bibnamefont {Casula}}, \bibinfo
  {author} {\bibfnamefont {M.~J.}\ \bibnamefont {Verstraete}}, \bibinfo
  {author} {\bibfnamefont {M.~A.~L.}\ \bibnamefont {Marques}}, \bibinfo
  {author} {\bibfnamefont {I.~V.}\ \bibnamefont {Tokatly}}, \ and\ \bibinfo
  {author} {\bibfnamefont {A.}~\bibnamefont {Rubio}},\ }\href {\doibase
  10.1103/PhysRevA.83.032503} {\bibfield  {journal} {\bibinfo  {journal} {Phys.
  Rev. A}\ }\textbf {\bibinfo {volume} {83}},\ \bibinfo {pages} {032503}
  (\bibinfo {year} {2011})}\BibitemShut {NoStop}%
\bibitem [{\citenamefont {Shulenburger}\ \emph {et~al.}(2009)\citenamefont
  {Shulenburger}, \citenamefont {Casula}, \citenamefont {Senatore},\ and\
  \citenamefont {Martin}}]{LDA_INT}%
  \BibitemOpen
  \bibfield  {author} {\bibinfo {author} {\bibfnamefont {L.}~\bibnamefont
  {Shulenburger}}, \bibinfo {author} {\bibfnamefont {M.}~\bibnamefont
  {Casula}}, \bibinfo {author} {\bibfnamefont {G.}~\bibnamefont {Senatore}}, \
  and\ \bibinfo {author} {\bibfnamefont {R.~M.}\ \bibnamefont {Martin}},\
  }\href {http://stacks.iop.org/1751-8121/42/i=21/a=214021} {\bibfield
  {journal} {\bibinfo  {journal} {J. Phys. A}\ }\textbf {\bibinfo {volume}
  {42}},\ \bibinfo {pages} {214021} (\bibinfo {year} {2009})}\BibitemShut
  {NoStop}%
\bibitem [{\citenamefont {Casula}\ \emph {et~al.}(2006)\citenamefont {Casula},
  \citenamefont {Sorella},\ and\ \citenamefont {Senatore}}]{LDA_INT2}%
  \BibitemOpen
  \bibfield  {author} {\bibinfo {author} {\bibfnamefont {M.}~\bibnamefont
  {Casula}}, \bibinfo {author} {\bibfnamefont {S.}~\bibnamefont {Sorella}}, \
  and\ \bibinfo {author} {\bibfnamefont {G.}~\bibnamefont {Senatore}},\ }\href
  {\doibase 10.1103/PhysRevB.74.245427} {\bibfield  {journal} {\bibinfo
  {journal} {Phys. Rev. B}\ }\textbf {\bibinfo {volume} {74}},\ \bibinfo
  {pages} {245427} (\bibinfo {year} {2006})}\BibitemShut {NoStop}%
\bibitem [{\citenamefont {Baker}\ \emph {et~al.}(2015)\citenamefont {Baker},
  \citenamefont {Stoudenmire}, \citenamefont {Wagner}, \citenamefont {Burke},\
  and\ \citenamefont {White}}]{LDA_DMRG}%
  \BibitemOpen
  \bibfield  {author} {\bibinfo {author} {\bibfnamefont {T.~E.}\ \bibnamefont
  {Baker}}, \bibinfo {author} {\bibfnamefont {E.~M.}\ \bibnamefont
  {Stoudenmire}}, \bibinfo {author} {\bibfnamefont {L.~O.}\ \bibnamefont
  {Wagner}}, \bibinfo {author} {\bibfnamefont {K.}~\bibnamefont {Burke}}, \
  and\ \bibinfo {author} {\bibfnamefont {S.~R.}\ \bibnamefont {White}},\ }\href
  {\doibase 10.1103/PhysRevB.91.235141} {\bibfield  {journal} {\bibinfo
  {journal} {Phys. Rev. B}\ }\textbf {\bibinfo {volume} {91}},\ \bibinfo
  {pages} {235141} (\bibinfo {year} {2015})}\BibitemShut {NoStop}%
\bibitem [{\citenamefont {Xianlong}\ \emph {et~al.}(2006)\citenamefont
  {Xianlong}, \citenamefont {Polini}, \citenamefont {Tosi}, \citenamefont
  {Campo}, \citenamefont {Capelle},\ and\ \citenamefont {Rigol}}]{LDA_AdHoc3}%
  \BibitemOpen
  \bibfield  {author} {\bibinfo {author} {\bibfnamefont {G.}~\bibnamefont
  {Xianlong}}, \bibinfo {author} {\bibfnamefont {M.}~\bibnamefont {Polini}},
  \bibinfo {author} {\bibfnamefont {M.~P.}\ \bibnamefont {Tosi}}, \bibinfo
  {author} {\bibfnamefont {V.~L.}\ \bibnamefont {Campo}}, \bibinfo {author}
  {\bibfnamefont {K.}~\bibnamefont {Capelle}}, \ and\ \bibinfo {author}
  {\bibfnamefont {M.}~\bibnamefont {Rigol}},\ }\href {\doibase
  10.1103/PhysRevB.73.165120} {\bibfield  {journal} {\bibinfo  {journal} {Phys.
  Rev. B}\ }\textbf {\bibinfo {volume} {73}},\ \bibinfo {pages} {165120}
  (\bibinfo {year} {2006})}\BibitemShut {NoStop}%
\bibitem [{\citenamefont {Hodgson}\ \emph {et~al.}(2013)\citenamefont
  {Hodgson}, \citenamefont {Ramsden}, \citenamefont {Chapman}, \citenamefont
  {Lillystone},\ and\ \citenamefont {Godby}}]{iDEA}%
  \BibitemOpen
  \bibfield  {author} {\bibinfo {author} {\bibfnamefont {M.~J.~P.}\
  \bibnamefont {Hodgson}}, \bibinfo {author} {\bibfnamefont {J.~D.}\
  \bibnamefont {Ramsden}}, \bibinfo {author} {\bibfnamefont {J.~B.~J.}\
  \bibnamefont {Chapman}}, \bibinfo {author} {\bibfnamefont {P.}~\bibnamefont
  {Lillystone}}, \ and\ \bibinfo {author} {\bibfnamefont {R.~W.}\ \bibnamefont
  {Godby}},\ }\href {\doibase 10.1103/PhysRevB.88.241102} {\bibfield  {journal}
  {\bibinfo  {journal} {Phys. Rev. B}\ }\textbf {\bibinfo {volume} {88}},\
  \bibinfo {pages} {241102} (\bibinfo {year} {2013})}\BibitemShut {NoStop}%
\bibitem [{\citenamefont {Gordon}\ \emph {et~al.}(2005)\citenamefont {Gordon},
  \citenamefont {Santra},\ and\ \citenamefont {K\"artner}}]{SoftenedCoulomb}%
  \BibitemOpen
  \bibfield  {author} {\bibinfo {author} {\bibfnamefont {A.}~\bibnamefont
  {Gordon}}, \bibinfo {author} {\bibfnamefont {R.}~\bibnamefont {Santra}}, \
  and\ \bibinfo {author} {\bibfnamefont {F.~X.}\ \bibnamefont {K\"artner}},\
  }\href {\doibase 10.1103/PhysRevA.72.063411} {\bibfield  {journal} {\bibinfo
  {journal} {Phys. Rev. A}\ }\textbf {\bibinfo {volume} {72}},\ \bibinfo
  {pages} {063411} (\bibinfo {year} {2005})}\BibitemShut {NoStop}%
\bibitem [{Note1()}]{Note1}%
  \BibitemOpen
  \bibinfo {note} {Spinless electrons obey the Pauli principle but are
  restricted to a single spin type. Systems of two or three spinless electrons
  exhibit features that would need a larger number of spin-half electrons to
  become apparent. For example, two spinless electrons experience the exchange
  effect, which is not the case for two spin-half electrons in an $S = 0$
  state. Furthermore, spinless KS electrons occupy a greater number of KS
  orbitals.}\BibitemShut {Stop}%
\bibitem [{Note2()}]{Note2}%
  \BibitemOpen
  \bibinfo {note} {We use Hartree atomic units: $m_{\protect \rm {e}} = \hbar =
  e = 4\pi \varepsilon _{0} = 1$}\BibitemShut {NoStop}%
\bibitem [{Note3()}]{Note3}%
  \BibitemOpen
  \bibinfo {note} {See Supplemental Material for the initial LDAs.}\BibitemShut
  {Stop}%
\bibitem [{Note4()}]{Note4}%
  \BibitemOpen
  \bibinfo {note} {See Supplemental Material for the errors.}\BibitemShut
  {Stop}%
\bibitem [{Note5()}]{Note5}%
  \BibitemOpen
  \bibinfo {note} {See Supplemental Material for the initial $1e$
  LDA.}\BibitemShut {Stop}%
\bibitem [{\citenamefont {Perdew}\ and\ \citenamefont
  {Wang}(1992)}]{PerdewWang}%
  \BibitemOpen
  \bibfield  {author} {\bibinfo {author} {\bibfnamefont {J.~P.}\ \bibnamefont
  {Perdew}}\ and\ \bibinfo {author} {\bibfnamefont {Y.}~\bibnamefont {Wang}},\
  }\href {\doibase 10.1103/PhysRevB.45.13244} {\bibfield  {journal} {\bibinfo
  {journal} {Phys. Rev. B}\ }\textbf {\bibinfo {volume} {45}},\ \bibinfo
  {pages} {13244} (\bibinfo {year} {1992})}\BibitemShut {NoStop}%
\bibitem [{\citenamefont {Cole}\ and\ \citenamefont
  {Perdew}(1982)}]{ColePerdew}%
  \BibitemOpen
  \bibfield  {author} {\bibinfo {author} {\bibfnamefont {L.~A.}\ \bibnamefont
  {Cole}}\ and\ \bibinfo {author} {\bibfnamefont {J.~P.}\ \bibnamefont
  {Perdew}},\ }\href {\doibase 10.1103/PhysRevA.25.1265} {\bibfield  {journal}
  {\bibinfo  {journal} {Phys. Rev. A}\ }\textbf {\bibinfo {volume} {25}},\
  \bibinfo {pages} {1265} (\bibinfo {year} {1982})}\BibitemShut {NoStop}%
\bibitem [{\citenamefont {Sun}\ \emph {et~al.}(2016)\citenamefont {Sun},
  \citenamefont {Perdew}, \citenamefont {Yang},\ and\ \citenamefont
  {Peng}}]{LDA_Finite}%
  \BibitemOpen
  \bibfield  {author} {\bibinfo {author} {\bibfnamefont {J.}~\bibnamefont
  {Sun}}, \bibinfo {author} {\bibfnamefont {J.~P.}\ \bibnamefont {Perdew}},
  \bibinfo {author} {\bibfnamefont {Z.}~\bibnamefont {Yang}}, \ and\ \bibinfo
  {author} {\bibfnamefont {H.}~\bibnamefont {Peng}},\ }\href {\doibase
  http://dx.doi.org/10.1063/1.4950845} {\bibfield  {journal} {\bibinfo
  {journal} {J. Chem. Phys}\ }\textbf {\bibinfo {volume} {144}},\ \bibinfo
  {eid} {191101} (\bibinfo {year} {2016}),\
  http://dx.doi.org/10.1063/1.4950845}\BibitemShut {NoStop}%
\bibitem [{Note6()}]{Note6}%
  \BibitemOpen
  \bibinfo {note} {This raises the question as to whether an LDA constructed
  from finite systems containing two spin-half electrons in the $S = 0$ state
  (fully spin-unpolarized) would be much closer to the LSDA ($\zeta =
  0$).}\BibitemShut {Stop}%
\bibitem [{\citenamefont {Hood}\ \emph {et~al.}(1997)\citenamefont {Hood},
  \citenamefont {Chou}, \citenamefont {Williamson}, \citenamefont {Rajagopal},
  \citenamefont {Needs},\ and\ \citenamefont {Foulkes}}]{QMC_Silicon}%
  \BibitemOpen
  \bibfield  {author} {\bibinfo {author} {\bibfnamefont {R.~Q.}\ \bibnamefont
  {Hood}}, \bibinfo {author} {\bibfnamefont {M.~Y.}\ \bibnamefont {Chou}},
  \bibinfo {author} {\bibfnamefont {A.~J.}\ \bibnamefont {Williamson}},
  \bibinfo {author} {\bibfnamefont {G.}~\bibnamefont {Rajagopal}}, \bibinfo
  {author} {\bibfnamefont {R.~J.}\ \bibnamefont {Needs}}, \ and\ \bibinfo
  {author} {\bibfnamefont {W.~M.~C.}\ \bibnamefont {Foulkes}},\ }\href
  {\doibase 10.1103/PhysRevLett.78.3350} {\bibfield  {journal} {\bibinfo
  {journal} {Phys. Rev. Lett.}\ }\textbf {\bibinfo {volume} {78}},\ \bibinfo
  {pages} {3350} (\bibinfo {year} {1997})}\BibitemShut {NoStop}%
\bibitem [{\citenamefont {Kim}\ \emph {et~al.}(2013)\citenamefont {Kim},
  \citenamefont {Sim},\ and\ \citenamefont {Burke}}]{Errors_Energy}%
  \BibitemOpen
  \bibfield  {author} {\bibinfo {author} {\bibfnamefont {M.-C.}\ \bibnamefont
  {Kim}}, \bibinfo {author} {\bibfnamefont {E.}~\bibnamefont {Sim}}, \ and\
  \bibinfo {author} {\bibfnamefont {K.}~\bibnamefont {Burke}},\ }\href
  {\doibase 10.1103/PhysRevLett.111.073003} {\bibfield  {journal} {\bibinfo
  {journal} {Phys. Rev. Lett.}\ }\textbf {\bibinfo {volume} {111}},\ \bibinfo
  {pages} {073003} (\bibinfo {year} {2013})}\BibitemShut {NoStop}%
\bibitem [{\citenamefont {{Durrant}}\ \emph {et~al.}(2015)\citenamefont
  {{Durrant}}, \citenamefont {{Hodgson}}, \citenamefont {{Ramsden}},\ and\
  \citenamefont {{Godby}}}]{ELF2}%
  \BibitemOpen
  \bibfield  {author} {\bibinfo {author} {\bibfnamefont {T.~R.}\ \bibnamefont
  {{Durrant}}}, \bibinfo {author} {\bibfnamefont {M.~J.~P.}\ \bibnamefont
  {{Hodgson}}}, \bibinfo {author} {\bibfnamefont {J.~D.}\ \bibnamefont
  {{Ramsden}}}, \ and\ \bibinfo {author} {\bibfnamefont {R.~W.}\ \bibnamefont
  {{Godby}}},\ }\href@noop {} {\bibfield  {journal} {\bibinfo  {journal} {ArXiv
  e-prints}\ } (\bibinfo {year} {2015})},\ \Eprint
  {http://arxiv.org/abs/1505.07687} {arXiv:1505.07687 [cond-mat.mes-hall]}
  \BibitemShut {NoStop}%
\bibitem [{\citenamefont {Hodgson}\ \emph {et~al.}(2014)\citenamefont
  {Hodgson}, \citenamefont {Ramsden}, \citenamefont {Durrant},\ and\
  \citenamefont {Godby}}]{iDEA2}%
  \BibitemOpen
  \bibfield  {author} {\bibinfo {author} {\bibfnamefont {M.~J.~P.}\
  \bibnamefont {Hodgson}}, \bibinfo {author} {\bibfnamefont {J.~D.}\
  \bibnamefont {Ramsden}}, \bibinfo {author} {\bibfnamefont {T.~R.}\
  \bibnamefont {Durrant}}, \ and\ \bibinfo {author} {\bibfnamefont {R.~W.}\
  \bibnamefont {Godby}},\ }\href {\doibase 10.1103/PhysRevB.90.241107}
  {\bibfield  {journal} {\bibinfo  {journal} {Phys. Rev. B}\ }\textbf {\bibinfo
  {volume} {90}},\ \bibinfo {pages} {241107} (\bibinfo {year}
  {2014})}\BibitemShut {NoStop}%
\bibitem [{\citenamefont {Dobson}(1991)}]{ELF}%
  \BibitemOpen
  \bibfield  {author} {\bibinfo {author} {\bibfnamefont {J.~F.}\ \bibnamefont
  {Dobson}},\ }\href@noop {} {\bibfield  {journal} {\bibinfo  {journal} {J.
  Chem. Phys}\ }\textbf {\bibinfo {volume} {94}},\ \bibinfo {pages} {4328}
  (\bibinfo {year} {1991})}\BibitemShut {NoStop}%
\bibitem [{\citenamefont {Becke}\ and\ \citenamefont {Edgecombe}(1990)}]{ELF3}%
  \BibitemOpen
  \bibfield  {author} {\bibinfo {author} {\bibfnamefont {A.~D.}\ \bibnamefont
  {Becke}}\ and\ \bibinfo {author} {\bibfnamefont {K.~E.}\ \bibnamefont
  {Edgecombe}},\ }\href@noop {} {\bibfield  {journal} {\bibinfo  {journal} {J.
  Chem. Phys.}\ }\textbf {\bibinfo {volume} {92}},\ \bibinfo {pages} {5397}
  (\bibinfo {year} {1990})}\BibitemShut {NoStop}%
\bibitem [{Note7()}]{Note7}%
  \BibitemOpen
  \bibinfo {note} {See Supplemental Material for the specific parameters of our
  test systems.}\BibitemShut {Stop}%
\bibitem [{Note8()}]{Note8}%
  \BibitemOpen
  \bibinfo {note} {The exact $V_{\protect \rm {xc}}$ is obtained up to an
  additive constant, which we choose so that $V_{\protect \rm {xc}}$
  asymptotically approaches zero as $|x| \rightarrow \infty $.}\BibitemShut
  {Stop}%
\bibitem [{Note9()}]{Note9}%
  \BibitemOpen
  \bibinfo {note} {The $V_{\protect \rm {xc}} = 0$ approximation (Hartree
  theory) provides a benchmark against which we test the LDAs' ability to
  describe the self-interaction correction.}\BibitemShut {Stop}%
\bibitem [{\citenamefont {Hellgren}\ and\ \citenamefont
  {Gross}(2012)}]{Steps4}%
  \BibitemOpen
  \bibfield  {author} {\bibinfo {author} {\bibfnamefont {M.}~\bibnamefont
  {Hellgren}}\ and\ \bibinfo {author} {\bibfnamefont {E.~K.~U.}\ \bibnamefont
  {Gross}},\ }\href {\doibase 10.1103/PhysRevA.85.022514} {\bibfield  {journal}
  {\bibinfo  {journal} {Phys. Rev. A}\ }\textbf {\bibinfo {volume} {85}},\
  \bibinfo {pages} {022514} (\bibinfo {year} {2012})}\BibitemShut {NoStop}%
\bibitem [{\citenamefont {Hodgson}\ \emph {et~al.}(2016)\citenamefont
  {Hodgson}, \citenamefont {Ramsden},\ and\ \citenamefont {Godby}}]{Steps}%
  \BibitemOpen
  \bibfield  {author} {\bibinfo {author} {\bibfnamefont {M.~J.~P.}\
  \bibnamefont {Hodgson}}, \bibinfo {author} {\bibfnamefont {J.~D.}\
  \bibnamefont {Ramsden}}, \ and\ \bibinfo {author} {\bibfnamefont {R.~W.}\
  \bibnamefont {Godby}},\ }\href {\doibase 10.1103/PhysRevB.93.155146}
  {\bibfield  {journal} {\bibinfo  {journal} {Phys. Rev. B}\ }\textbf {\bibinfo
  {volume} {93}},\ \bibinfo {pages} {155146} (\bibinfo {year}
  {2016})}\BibitemShut {NoStop}%
\bibitem [{\citenamefont {Lein}\ and\ \citenamefont {K\"ummel}(2005)}]{Steps2}%
  \BibitemOpen
  \bibfield  {author} {\bibinfo {author} {\bibfnamefont {M.}~\bibnamefont
  {Lein}}\ and\ \bibinfo {author} {\bibfnamefont {S.}~\bibnamefont
  {K\"ummel}},\ }\href {\doibase 10.1103/PhysRevLett.94.143003} {\bibfield
  {journal} {\bibinfo  {journal} {Phys. Rev. Lett.}\ }\textbf {\bibinfo
  {volume} {94}},\ \bibinfo {pages} {143003} (\bibinfo {year}
  {2005})}\BibitemShut {NoStop}%
\bibitem [{\citenamefont {Vieira}\ \emph {et~al.}(2009)\citenamefont {Vieira},
  \citenamefont {Capelle},\ and\ \citenamefont {Ullrich}}]{Steps3}%
  \BibitemOpen
  \bibfield  {author} {\bibinfo {author} {\bibfnamefont {D.}~\bibnamefont
  {Vieira}}, \bibinfo {author} {\bibfnamefont {K.}~\bibnamefont {Capelle}}, \
  and\ \bibinfo {author} {\bibfnamefont {C.~A.}\ \bibnamefont {Ullrich}},\
  }\href@noop {} {\bibfield  {journal} {\bibinfo  {journal} {Physical Chemistry
  Chemical Physics}\ }\textbf {\bibinfo {volume} {11}},\ \bibinfo {pages}
  {4647} (\bibinfo {year} {2009})}\BibitemShut {NoStop}%
\end{thebibliography}%
\end{document}